\documentclass[amsmath,a4paper,twocolumn,superscriptaddress,floatfix,aps,pra,10pt]{revtex4-1} %
\usepackage{amssymb}
\usepackage{amsmath} 
\usepackage{graphicx}
\usepackage{amsfonts}
\usepackage{color}
\usepackage{bbm}
\usepackage{tcolorbox}

\tcbuselibrary{theorems}

\tcbset{colback=green!5,colframe=green!50!blue,
fonttitle=\bfseries, float=htb}

\newtcolorbox[blend into=figures]{boxfigure}[3][]
{ float*=ht,width=\textwidth,lower separated=false, center upper, 
title={#2},label= fig:#3,#1}

\newtcolorbox[blend into=figures]{smallboxfigure}[3][]
{float*=ht,lower separated=false, center upper, 
title={#2}, label= fig:#3 ,#1}

\newtcolorbox[auto counter]{example}[3][]
{float*=ht,title=Example ~\thetcbcounter: #2,label= ex:#3 ,#1}

\newtcolorbox[use counter from=example]{bigexample}[3][]
{ float*=ht,width=\textwidth, title=Example ~\thetcbcounter: #2,every float=\centering,label= ex:#3,#1}

\newcommand{\bra}[1]{\ensuremath{\left\langle #1\right|}}
\newcommand{\ket}[1]{\ensuremath{\left|#1\right\rangle}}
\newcommand{\proj}[1]{\ensuremath{\left|#1\right\rangle\left\langle #1\right|}}

\newcommand{\norm}[1]{\left\| #1\right\|}

\newcommand{\bea}{\begin{eqnarray}}
\newcommand{\eea}{\end{eqnarray}}

\newcommand{\id}{\mathbbm{1}}

\newcommand{\rc}{\mathrm{C}}
\newcommand{\rr}{\mathrm{R}}
\newcommand{\rh}{\mathrm{H}}
\newcommand{\rst}{\mathrm{st}}
\newcommand{\st}{\mathrm{st}} 
\newcommand{\rint}{\mathrm{int}}

\newcommand{\ba}{\begin{eqnarray}}
\newcommand{\ea}{\end{eqnarray}}

\DeclareMathOperator{\Tr}{Tr}
\DeclareMathOperator{\tr}{Tr}

\newcommand{\integral}[3]{{\int^{#2}_{#3} \mathrm{d}{#1} \;}} 

\newcommand{\pure}[1]{\proj{#1}}
\newcommand{\avg}[1]{\langle #1 \rangle}

\newcommand{\cT}{\mathcal{T}}
\newcommand{\hmax}{H_{\max}}

\newcommand{\COP}{\mathrm{COP}}
\input{Qcircuit.tex}
\usepackage[pdftex,breaklinks,colorlinks]{hyperref}

\begin{document}

\title{The role of quantum information in thermodynamics --- a topical review}

\author{John Goold}
\affiliation{The Abdus Salam International Centre for Theoretical Physics (ICTP), Trieste, Italy}
\author{Marcus Huber}
\affiliation{Group of Applied Physics, University of Geneva, 1211 Geneva 4, Switzerland}
\affiliation{Universitat Autonoma de Barcelona, 08193 Bellaterra, Barcelona, Spain}
\affiliation{ICFO-Institut de Ciencies Fotoniques, The Barcelona Institute of Science and Technology, 08860 Castelldefels (Barcelona), Spain}
\author{Arnau Riera}
\affiliation{ICFO-Institut de Ciencies Fotoniques, The Barcelona Institute of Science and Technology, 08860 Castelldefels (Barcelona), Spain}
\author{L\'idia del Rio}
\affiliation{H. H. Wills Physics Laboratory, University of Bristol, Bristol BS8 1TL, United Kingdom}
\author{Paul Skrzypczyk}
\affiliation{ICFO-Institut de Ciencies Fotoniques, The Barcelona Institute of Science and Technology, 08860 Castelldefels (Barcelona), Spain}
\affiliation{H. H. Wills Physics Laboratory, University of Bristol, Bristol BS8 1TL, United Kingdom}

\date{\today}

\begin{abstract}
This topical review article gives an overview of the interplay between 
 quantum information theory and thermodynamics of quantum systems. We focus on several trending topics including the foundations of statistical mechanics, resource theories, entanglement in thermodynamic settings, fluctuation theorems and thermal machines. 
 This is not a comprehensive review of the diverse field of quantum thermodynamics; rather, it is a convenient entry point for the thermo-curious information theorist. Furthermore this review should facilitate the unification and understanding of different interdisciplinary approaches emerging in research groups around the world.
\end{abstract}

\maketitle

\tableofcontents

\newpage

\section{Introduction} 
\label{sec:introduction}

\begin{boxfigure}{Maxwell's demon}{demon}

\includegraphics[width=0.5 \textwidth]{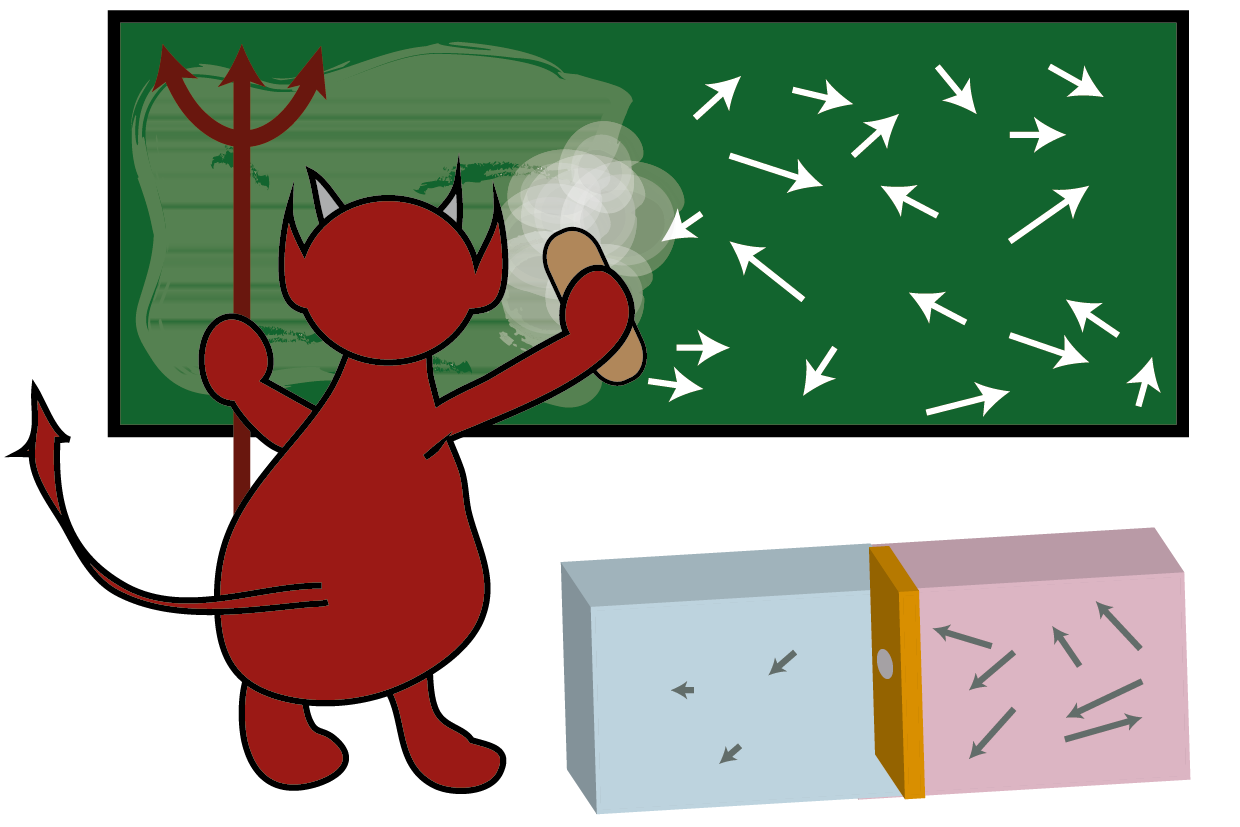}
\tcblower

 An early puzzle in thermodynamics: imagine a box filled with a gas, with a partition in the middle. An agent (the demon) who can observe the microscopic details of the gas particles, controls a small gate in the partition, selectively opening it to let slow particles flow to the left and fast ones to the right. This creates a temperature differential between the two sides. The demon can exploit this difference to extract work, by letting the hot gas on the right expand. This apparent contradiction with the second law of thermodynamics can be easily explained from an information-theoretical viewpoint. The demon had access to much more information than the standard observer assumed in the derivation of traditional thermodynamics, who can only read a few macroscopic parameters of a gas  and assumes a uniform distribution over all compatible micro-states. Therefore, it seems natural that the demon may extract more work than predicted by standard thermodynamics|and this insight motivates the need for thinking of thermodynamics as a subjective resource theory, and extending it to the quantum regime. 
In the larger picture,  Bennett showed that the amount of work needed to erase the demon's memory at the end of the procedure (or equivalently, to prepare the memory  to store the necessary information on the particles in the beginning) precisely makes up for the work extracted
\cite{Bennett1982}. For  reviews, see \cite{LeffRex90,Plenio2001,LeffRex02,Maruyama2009,Parrondo2015}.

\end{boxfigure}

If physical theories were people, thermodynamics would be the village witch. Over the course of  three centuries, she smiled quietly as other theories rose and withered, surviving major revolutions in physics, like the advent of general relativity and quantum mechanics. The other theories find her somewhat odd,  somehow \emph{different} in nature from the rest, yet everyone comes to her for advice, and no-one dares to  contradict her. Einstein, for instance, called her ``the only physical theory of universal content, which I am convinced, that within the framework of applicability of its basic concepts will never be overthrown.''

Her power and resilience lay mostly on her frank intentions: thermodynamics has never claimed to be a means to  understand the mysteries of the natural world, but rather a path towards efficient exploitation of said world. She tells us how to make the most of some resources, like a hot gas or a magnetized metal, to achieve specific goals, be them moving a train or formatting a hard drive. Her universality  comes from the fact that she does not try to understand the microscopic details of particular systems. Instead, she only cares to identify which operations are easy and hard to implement in those systems, and which resources are freely available to an experimenter,  in order to quantify the cost of state transformations.
Although it may stand out within physics, this operational approach can be found in branches of computer science, economics and mathematics, and it plays a central role in quantum information theory|which is arguably why quantum information, a toddler among physical theories, is bringing so much to thermodynamics. 

In the early twentieth century, information theory was constructed as the epitome of detachment from physics \cite{Shannon1948}. Its basic premise was that we could think of information independently of its physical support: a message in a bottle, a bit string and a sensitive phone call could all be treated in the same way.
This level of abstraction was not originally conceived for its elegance; rather, it emerged as the natural way to address very earthly questions, such as  ``can I reliably send a message through a noisy line?'' and ``how much space do I need to store a picture?''. 
In trying to quantify the resources required by those tasks (for example, the number of uses of the noisy channel, or of memory bits),  it soon became clear that the relevant quantities were  variations of what is now generally known as entropy \cite{TC06}. 
\emph{Entropy measures} quantify our uncertainty about events: they can tell us how likely we are to guess the outcome of a coin toss, or the content of a message, given some side knowledge we might have. As such, they depend only on probability distributions over those events, and not on their actual content (when computing the odds, is does not matter whether they apply to a coin toss or to a horse race). 
Information theory has been greatly successful in this approach, and is used in fields from file compression to practical cryptography and channel coding \cite{TC06}.

But as it turned out, not all information was created equal. 
If we zoom in and try to encode information in the tiniest support possible, say the spin of an electron,  we face some of the perplexing aspects of quantum physics: we can write in any real number, but it is only possible to read one bit out, we cannot copy information, and we find correlations that cannot be explained by local theories. In short, we could not simply apply the old information theory to tasks involving quantum particles, and the scattered study of quirky quantum effects soon evolved into the fully-fledged discipline of quantum information theory \cite{nielsenchuang}. Today we see quantum theory as a generalization of classical probability theory, with density matrices replacing probability distributions, measurements taking the place of events, and quantum entropy measures to characterize operational tasks \cite{Wilde2013}.

\begin{boxfigure}{The thermodynamic origin of the von Neumann entropy}{vonNeumannEntropy}

\includegraphics[width=0.5 \textwidth]{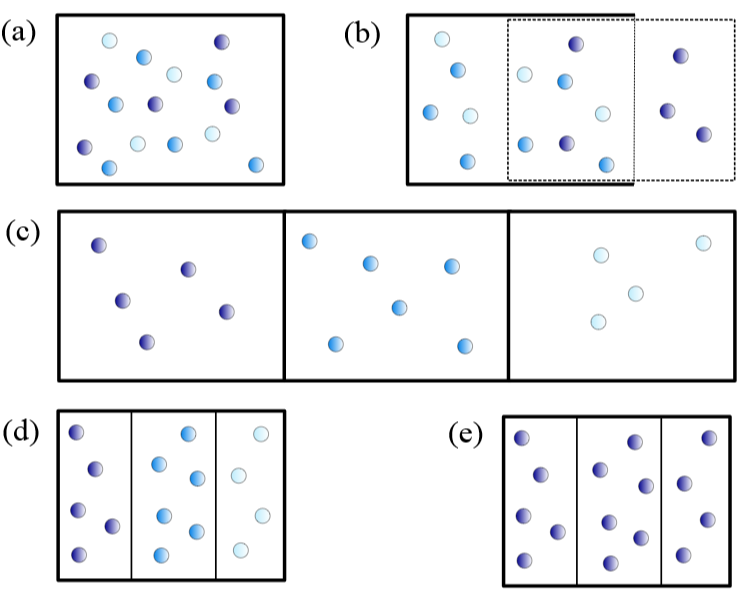}
\tcblower

In 1932,  von Neumann designed this thought experiment to determine the entropy of a density operator $\rho$ \cite{vonNeumann1955}. The experiment accounts for the work cost of erasing the state of a gas of  $n$ atoms, initially in an ensemble  $\rho^{\otimes n}$, 
with $\rho=\sum_k p_k\proj{\phi_k}$, by transforming it into a pure state $\ket{\phi_1}^{\otimes n}$ 
by means of a \emph{reversible} process.

It consists of 3 steps:
1. \emph{Separation of the species}: the atoms in different states $\ket{\phi_1}$,\ldots,$\ket{\phi_m}$ inside a box of volume $V$ are separated in different boxes of the same volume $V$ by means of semi-permeable walls (from a to b and finally c). Note that no work has been done and no heat has been exchanged.
2. \emph{Compression}: every gas $\ket{\phi_k}$ is isothermally compressed to a volume $V_k=p_k V$ (from c to d). The mechanical work done in that process is $W_k=n p_k \ln (V_k/V)= p_k \ln p_k$. The total entropy increase per particle of that process is
$\Delta S= \sum_k p_k \ln p_k$.
3. \emph{Unitary transformation}: every gas is put in
the $\ket{\phi_1}$ state by applying different unitary transformations $\ket{\phi_k}\to\ket{\phi_1}$, which are taken for free (from d to e). 
As the entropy of the final state is zero, the entropy of the initial ensemble reads
$ S(\rho)= -\Tr(\rho\ln \rho)$.

Historically, it is remarkable that the Shannon entropy, which can be seen as
a particular case of the von Neumann entropy for classical ensembles, was not introduced until 1948 \cite{Shannon1948}, and Landauer's principle was proposed only in 1961 \cite{Landauer1961}.
\end{boxfigure}

While quantum information theory has helped us understand the nature of the quantum world, its practical applications are not as well spread as for its classical counterpart.  Technology is simply not there yet|not at the point where we may craft, transport and preserve all the quantum states necessary in a large scale. These technical limitations, together with a desire to pin down exactly what makes quantum special, gave rise to resource theories within quantum information, for instance theories of entanglement \cite{Horodecki2009}. There, the rough premise is that entangled states are useful for many interesting tasks (like secret key sharing), but distributing entanglement over two or more agents by transporting quantum particles over a distance is hard, as there are always losses in the process \cite{Bauml2015}. Therefore, all entangled states become a precious resource, and we study how to distill entanglement from them using only a set of allowed operations, which are deemed to be easier to implement|most notoriously, local operations and classical communication \cite{Bennett1996}. 

Other resource theories started to emerge within quantum information|purity and asymmetry have also been framed as resources under different sets of constraints|and this way of thinking quickly spread among the quantum information community (see Ref.~\cite{Coecke2014} for a review). As many of its members have a background in physics and an appetite for abstraction, it was a natural step for them to approach thermodynamics with such a framework in mind. Their results strengthen thermodynamics, not only by extending her range of applicability to small quantum systems, but also by revisiting her fundamental principles. The resource theory approach to thermodynamics is reviewed in Section~\ref{sec:resource_theories}.

Each resource theory explores the limitations imposed by one specific physical constraint, like locality or energy conservation. In a realistic setting we could be bound to several of these constraints, a natural case that can be modelled by combining different resource theories, thus restricting the set of allowed operations.
In Section~\ref{sec:entanglement} we review and discuss attempts to combine thermodynamic and locality constraints. In particular, we look at the role of entanglement resources in thermodynamic tasks, thermodynamic witnesses of non-classicality, and entanglement witnesses in phase transitions.

Information theory also shed light on fundamental issues in statistical mechanics - the mathematical backbone of thermodynamics. Perhaps one of the earliest significant contributions is the maximal entropy principle introduced by Jaynes \cite{Jaynes1957,Jaynes1957a}. In these seminal works Jaynes addresses the issue of justifying the methods of statistical mechanics from microscopic mechanical laws (classical or quantum) using tools from information theory. In fact, deriving statistical mechanics, and hence thermodynamics from quantum mechanics is almost as old as quantum mechanics itself starting with the work of von Neumann \cite{vonNeumann1955, vonNeumann2010}. This is still very much and ongoing and active research area and in recent years has received significant attention from the quantum information community. The most significant contributions are reviewed in Section~\ref{sec:foundations}.

In the past twenty  years, the field of non-equilibrium statistical mechanics has seen a rapid development in the treatments of driven classical and quantum systems beyond the linear response regime. This has culminated in the discovery of various fluctuation theorems which relate equilibrium thermodynamic quantities to non-equilibrium ones, and led to a revision on how we understand the thermodynamics of systems far from equilibrium \cite{Jarzynski2011,Esposito2009,Campisi2011,Seifert2012,Hanggi2015}. Although this approach is relatively recent from a statistical physics perspective, a cross-fertilisation with concepts ubiquitous in quantum information theory has already started, including phase estimation techniques for extraction of work and heat statistics and feedback fluctuation theorems for Maxwell's demons. In Section~\ref{sec:fluctuations} we identify these existing relationships and review areas where more overlap could be developed.

As ideas and concepts emerge and develop it is not surprising that quantum information theorists have started to turn towards the pragmatic goal of describing the advantages and disadvantages of machines which operate at and below the quantum threshold. Although ideas relating quantum engines have been around now for a surprising long time \cite{Scovil1959,Geusic1967,Alicki1979} - questions pertaining to the intrinsic quantumness in the functioning of such machines have been raised using the tools of quantum information theory only relatively recently. We review progress along these lines in Section~\ref{sec:thermal_machines}.
In summary, we will review landmark and recent articles in quantum thermodynamics, discuss different approaches and models, and peek into future directions of the field.  


\section*{Scope and other reviews}
\label{sec:reviews}

This review focuses on landmark and recent articles in the field of quantum thermodynamics with a special emphasis on contributions from quantum information theory. We place emphasis on current trending topics, discuss different approaches and models and peek into the future directions of the field. As the review is ``topical'', we intend to focus our attention on the interplay of quantum information and thermodynamics and we have written such that interested readers from different communities will get a detailed overview of how their respective techniques have been successfully applied to provide a deeper understanding of the field.

As the vastness of possible topics could easily exceed the scope of a topical review, we refer to other review articles and books concerning questions that have already been covered by other authors:

\begin{itemize}

\item  \textbf{Equilibration and thermalization.}
Recovering statistical mechanics from the unitary evolution of a closed quantum system is an issue which is almost as old as quantum mechanics itself. This topic, far from being an academic issue, has seen an unprecedented revival of interest due mainly to advances in experimental ultra-cold atoms. We discus the topic in Section~\ref{sec:foundations}, from a quantum information perspective. This topic is more extensively reviewed in Ref.~\cite{Gogolin2015}. For readers interested in this topic from a condensed matter perspective we recommend the review \cite{Polkovnikov2011} and the special issue \cite{Daley2014} for more recent developments.

\item  \textbf{Thermal machines.} As mentioned in the introduction viewing engine cycles from a fully quantum mechanical perspective is also not a new topic \cite{Scovil1959,Geusic1967,Alicki1979}. Many results on quantum engines exist which are not directly related to quantum information processing we exclude them from Section~\ref{sec:thermal_machines} and the interested reader may learn more in Refs.~\cite{Kosloff2013a,Kosloff2014,Gelbwaser2015a}.

\item  \textbf{Maxwell's demon and Landauer's principle.} Almost as old as thermodynamics itself is the Maxwell's demon paradox, briefly introduced in Figure~\ref{fig:demon} and Example~\ref{ex:erasure_intro}. The demon paradox inspired the seminal work of Szilard to reformulate the demon as a binary decision problem \cite{Szilard}. The resolution of Maxwell demon paradox by Landauer cements the relationship between the physical and information theoretical worlds. This demon has been extensively investigated from both a quantum and classical perspective in Refs.~\cite{LeffRex90,Plenio2001,LeffRex02,Maruyama2009, Parrondo2015}.

\item  \textbf{Quantum thermodynamics.}
The 2009 book  \cite{mahlerbook} covers a range of topics regarding the emergence of thermodynamic behaviour in composite quantum systems. 

\item  \textbf{Entanglement and phase transitions in condensed matter.}
Entanglement is frequently used as an indicator of quantum phase transitions in condensed matter systems. We do not cover this particular setting but the interested reader may find a comprehensive review in Ref.~\cite{Amico2008a}.

\item \textbf{Resource theories.} Examples and common features of resource theories (beyond quantum information theory) are discussed in Ref.~\cite{Fritz2015}. In particular, different approaches to general frameworks are discussed in Section 10 of that work.

\item \textbf{Experimental implementations.} Experiments with  demons, thermal engines and work extraction are discussed in more depth in the  perspective article \cite{Millen2015}. 

\end{itemize}


\begin{tcolorbox}[title=Definitions and notation]

Conventions followed unless otherwise stated:

{\bf States.} Discrete Hilbert spaces $\mathbbm{C}^d$. States $\rho$ are represented by Hermitian matrices ($\mathrm{Tr}(\rho)=1$ and $\rho\geq 0$). Subsystems are denoted by Roman subscripts,  $\rho_A:=\mathrm{Tr}_B(\rho_{AB})$.

{\bf Entropy.} Von Neumann entropy with  base $2$ logarithm, $S(\rho)= - \tr(\rho \ \log_2 (\rho))$.

{\bf Mutual information.} Measures  correlations,
$I(A:B)_\rho:=S(\rho_A)+S(\rho_B)-S(\rho_{AB})$.

{\bf Energy.} Hamiltonian $H$, average energy $\langle H \rangle_\rho=\tr(\rho H)$, eigenvalues $\{E_k\}_k$, eigenvectors $\{\ket{E_k}\}_{k}$, or $\{\ket{E_k^i }\}_{k,i}$ if there are degeneracies, with energy projectors $\Pi_k = \sum_i \pure{E_k^i}$.

{\bf Thermal states.} Gibbs state  $\tau(\beta)=\frac{e^{-\beta H}}{\mathcal{Z}}$, with partition function $\mathcal{Z}=\mathrm{Tr}(e^{-\beta H})$ and inverse temperature $\beta:=\frac{1}{k_B T}$.

{\bf Free energy.} $F_\beta(\rho):=\langle H \rangle_\rho-\frac{1}{\ln(2)\beta} S(\rho)$.

{\bf Linbladian.} $\mathcal{L}(\rho)$  generates Markovian, time-homogeneous, non-unitary dynamics.

\end{tcolorbox}

\section{Foundations of statistical mechanics}
\label{sec:foundations}


At first sight, thermodynamics and quantum theory are incompatible. 
While thermodynamics and statistical mechanics state that the entropy of the universe as a whole is a monotonically increasing quantity, according to quantum theory the entropy of the universe is constant since it evolves unitarily.
This leads us to the question of to which extent the methods of statistical physics can be justified from the microscopic theory of quantum mechanics and both theories can be made compatible. 
Unlike classical mechanics, quantum mechanics has a way to circumvent this paradox: \emph{entanglement}. 
We observe entropy to grow in physical systems because they are entangled
with the rest of the universe.  
In this section we review the progress made on this topic in recent years which show that equilibration and thermalization are intrinsic to quantum theory.

\subsection{Equal a priori probabilities postulate as a consequence of typicality in Hilbert spaces}
\label{sec:typicality}
Let us consider a closed system that evolves in time restricted to some global constraint. 
The principle of \emph{equal a priori probabilities} states that, at equilibrium, the system is equally likely to be found in any of its accessible states. This assumption lies at the heart of statistical mechanics since it allows for the computation of equilibrium expectation values by performing averages on the phase space. However, there is
no reason in the laws of mechanics (and quantum mechanics) to suggest that the system explores its set of  accessible states uniformly. Therefore, the \emph{equal a priori probabilities} principle has to be put in by hand.

One of the main insights from the field of quantum information theory to statistical mechanics is the substitution of the \emph{Equal a priori probabilities postulate} by the use of \emph{typicality} arguments \cite{Popescu2006,Goldstein2006}.
To be more precise, let us consider a quantum system described by a Hilbert space $\mathcal{H}_S\otimes\mathcal{H}_B$
where $\mathcal{H}_S$ contains the degrees of freedom that are experimentally accessible and $\mathcal{H}_B$ the ones that are not.
In practice, we think of $S$ as a subsystem that we can access, and $B$ as its environment
(sometimes called the bath). 
Concerning the global constraint, in classical mechanics, it is defined by the constants of motion of the system.
In quantum mechanics, we model the restriction as a subspace $\mathcal{H}_R \subseteq \mathcal{H}_S\otimes\mathcal{H}_B$.
Let us denote by $d_R$, $d_S$ and $d_B$ the dimensions of the Hilbert spaces $\mathcal{H}_R$,  $\mathcal{H}_S$ and $\mathcal{H}_B$ respectively.

The equal a priori probability principle would describe the equilibrium state as
\begin{equation}
\varepsilon_R = \frac{\id_R}{d_R} \, ,
\end{equation}
and would imply the state of the subsystem $S$ to be
\begin{equation}
\Omega_S = \Tr_B \varepsilon_R \, .
\end{equation}
In Ref.~\cite{Popescu2006} it is shown that, if we look only at the subsystem $S$, most of the states
in $\mathcal{H}_R$ are indistinguishable from the equal a priori probability state,
i.~e.~for most $\ket{\psi}\in\mathcal{H}_R$, $\Tr_B \proj{\psi}\approx \Omega_S$.
More explicitly, if $\ket{\psi}$ is randomly chosen  in $\mathcal{H}_R$ according to the uniform distribution
given by the Haar measure, 
then the probability that $\Tr_B \proj{\psi}$ can be distinguished from $\Omega_S$ 
decreases exponentially with the dimension of $\mathcal{H}_R$, $d_R$
\begin{align}
 {\rm Prob}\big[ \norm{ \Tr_B(\proj{\psi}) - \Omega_S}_1 \geq d_R^{-1/3} \big] \leq & 2 \exp\left( - C d_R^{1/3} \right) \, ,
  \label{eq:typical_thermalization_haar}
\end{align}
where $C$ is a constant and $\norm{\cdot}_1$ is the trace norm.
The trace norm $\norm{\rho - \sigma}_1$ measures the physical distinguishability between the states $\rho$ and $\sigma$
in the sense that a $\norm{\rho - \sigma}_1=\sup_{O\le \id}| \Tr(O\rho)-\Tr(O\sigma)|$, where the maximization is
made over all the observables $O$ with operator norm bounded by 1.
The proof of Eq.~(\ref{eq:typical_thermalization_haar}) relies upon concentration of measure and in particular on \emph{Levy's Lemma} 
(see Ref.~\cite{Popescu2006} for details).
Let us mention that ideas in this spirit can be already found in S.~Lloyd's Ph.D.\ thesis \cite{Lloyd-phd} published in 1991. In particular,
he presents bounds on how the expectation values of a fixed operator taken over random pure states of a restricted subspace fluctuate.

The weakness of the previous result lies in that the use of typicality is made in the whole  subspace $\mathcal{H}_R$
and, as we will justify next, this is not a physical assumption. 
In nature, Hamiltonians have local interactions and systems evolve for times that are much smaller than the age of the universe. 
Most states in the Hilbert space simply cannot be generated
by evolving an initial product state under 
an arbitrary time-dependent local Hamiltonian in a time that scales polynomially in the system size  \cite{Poulin2011}. 
Therefore, sampling uniformly from the whole Hilbert space is not physically meaningful. 
There has been a strong effort to generalize the concept of typicality for different sets of states \cite{Garnerone2010,Brandao2012designs,Hamma2012}.

The first ``realistic'' set of states in which typicality was studied was the set of
\emph{matrix product states} (MPS)  \cite{Affleck1988, Vidal2004}. 
These type of states have been proven to describe ground states of one-dimensional gapped Hamiltonians.
They are characterized by the rank of a bipartition of the
state. This parameter quantifies the maximum entanglement between partitions of an MPS.
The MPSs with fixed rank form a set of states with an efficient classical representation
(they only require polynomial resources in the number of particles).  
In Ref.~\cite{Garnerone2010}, it is proven that typicality occurs for the expectation value of
subsystems observables when the rank of the MPS scales polynomially with the
size of the system with a power greater than 2.

Another set recently considered in the literature has been the so called set 
of \emph{physical states} which consists
of all states that can be produced by evolving an initial product state with a local Hamiltonian for a time polynomial in the number of particles $n$. By Trotter decomposing the Hamiltonian, such a set can be proven to be equivalent to the set of local \emph{random quantum circuits}, that is, quantum circuits of qubits composed of polynomially many nearest neighbour two-qubit gates \cite{Poulin2011}.
In Ref.~\cite{Harrow2009a}, it was shown that the local random quantum circuits form an approximate unitary 2-design, i.~e.~that random circuits of only polynomial length will approximate the first and second moments of the Haar distribution.
In Ref.~\cite{Brandao2012designs} the previous work was extended to
poly($n$)-designs.
Finally, let us mention that the entanglement properties of typical \emph{physical states} were studied in Ref.~\cite{Hamma2012}. 

Let us mention that $k$-designs also appear naturally in the context of \emph{decoupling theorems}
in which a the subsystem $S$ undergoes a physical evolution
separated from the environment $B$, and one wonders under what conditions this evolution
destroys all initial correlations between $S$ and $B$. In particular, 
in Ref.~\cite{Szehr2013} it is shown that almost-2-designs decouple the subsystem $S$ from $B$
independently of $B$'s size.

Another objection against typicality is that there are many physically interesting systems, e.~g.\ integrable models, which, although
their initial state belongs to a certain restricted subspace $\mathcal{H}_R$, their expectation values
differ from the completely mixed state in $R$, $\varepsilon_R$, as expected from typicality arguments.
This is a consequence of the fact that their trajectories in the Hilbert subspace $\mathcal{H}_R$ don't lie for the
overwhelming majority of times on \emph{generic} states (see Fig.~\ref{fig:untypical-trajectory}).
Hence, in practice, statements on equilibration and thermalization will depend on the dynamical properties
of every system, that is, on their Hamiltonian. 
This leads us to the notion of \emph{dynamical typicality}.
In contrast to the \emph{kinematic typicality} presented in this section, where an ensemble has been
defined by all the states that belong to a certain subspace, in \emph{dynamical typicality}
the ensemble is defined by all states that share the same constants of motion given a Hamiltonian $H$ and an initial state $\ket{\psi(0)}$. 
Studying whether typicality also holds in such a set will be precisely the problem addressed in the next section.

\begin{boxfigure}{Typical and untypical trajectories}{untypical-trajectory}

\includegraphics[width=0.5\textwidth]{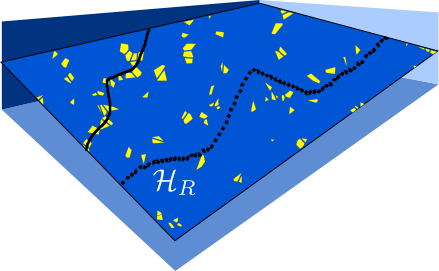}

\tcblower

 Scheme of the restricted subspace $\mathcal{H}_R$ with its untypical states forming little islands coloured in yellow. The left trajectory (dashed line) passes mostly on typical states while the right trajectory (solid line) has a non-negligible support on states that are not typical.

\end{boxfigure}


\subsection{Equilibration. Maximum entropy principle from quantum dynamics}
\label{sec:equilibration}


In this context of deriving thermodynamics from quantum mechanics the first problem that needs to be addressed is \emph{equilibration}, that is, understand how the reversible unitary dynamics of quantum mechanics make systems equilibrate and evolve towards a certain state where they remain thereafter.

Because of the unitary dynamics, equilibration is only possible if the set of observables is restricted.
In this spirit, a set of sufficient conditions for equilibration towards the time averaged state has been presented for local observables \cite{Linden2009,Short2012} and observables of finite precision \cite{Reimann2008,Reimann2012}.
The two approaches are proven to be equivalent in Ref.~\cite{Short2011}
and it is remarkable that the conditions given are weak 
and naturally fulfilled in realistic situations.

For simplicity, let us here focus on equilibration of subsystems
and, as above, identify in the total system a subsystem $S$ and its environment $B$.
The dynamics of the total system are governed by the Hamiltonian $H$ with eigenvalues $\{E_k\}_k$ and eigenvectors $\{\ket{E_k} \}_k$.
This leads to the time evolution 
$\ket{\psi(t)}= e^{-i H t} \ \ket{\psi(0)}$ 
and the reduced state of $S$ is $\rho_S(t)=\Tr_B \rho(t)$  with 
$\rho(t) = |\psi(t)\rangle \langle\psi(t)|$.

If equilibration happens, then it happens towards the time averaged state i.~e.\ 
$\omega_S := \Tr_B \omega$ with
\begin{equation}\label{eq:equilibrium-state}
	 \omega = \lim_{T \rightarrow \infty} \frac{1}{T} \int_0^{T} \rho(t) \mathrm{d}t 
	 = \sum_k  P_k \rho(0) P_k 
\end{equation}
with $P_k$ the projectors onto the Hamiltonian eigenspaces.
The time averaged state is the initial state dephased in the Hamiltonian eigenbasis.
For this reason it is also called \emph{diagonal ensemble}.

In Ref.~\cite{Linden2009}, a notion of equilibration
is introduced by means of the average distance (in time) of the subsystem $\rho_S(t)$ from equilibrium.
A subsystem $S$ is said to equilibrate if
\begin{equation}
\langle \norm{\rho_S(t)-\omega_S}_1\rangle_t:=
\lim_{T\to \infty }\frac{1}{T}\int_0^T \mathrm{d} t \norm{\rho_S(t)-\omega_S}_1
\ll 1\, ,
\end{equation}
where $\norm{\rho_S(t)-\omega_S}_1$ is the trace distance. 
If this average trace distance can be proven to be small, then
the subsystem $S$ is indistinguishable from being at equilibrium for almost all times.

Equilibration as a genuine property of quantum mechanics is shown in Ref.~\cite{Linden2009}
by precisely proving that this average distance is typically small. More concretely, if the Hamiltonian that dictates the
evolution of the system has \emph{non-degenerate gaps} i.~e.~all the gaps of the Hamiltonian are different (an assumption which we will comment on below), 
then the average distance from equilibrium is bounded by
\begin{equation}\label{eq:equilibration}
\langle \norm{\rho_S(t)-\omega_S}_1\rangle_t \leq \sqrt{\frac{d_S}{d^{\textrm{eff}}(\omega^B)}}
\leq \sqrt{\frac{d_S^2}{d^{\textrm{eff}}(\omega)}}\, ,
\end{equation}
where $d^{\textrm{eff}}(\rho):=1/\Tr(\rho^2)$ is the \emph{effective dimension} of $\rho$
and $\omega_B=\Tr_S \omega$.
Roughly speaking, the effective dimension of a state tells us how many eigenstates of the Hamiltonian support
such state. It can also be related to the 2-Renyi entanglement entropy by $S_2(\rho)=\log d^{\textrm{eff}}(\rho)$. Hence, equation (\ref{eq:equilibration}) guarantees equilibration for Hamiltonians with non-degenerate energy gaps
as long as the initial state is spread over many different energies.

Although the condition of having non-degenerate gaps may look very restrictive at first sight, 
note that Hamiltonians that do not fulfil it form a set of zero measure in the set of Hamiltonians,
since any arbitrarily weak perturbation breaks the degeneracy of the gaps. 
In Ref.~\cite{Short2012}, the non-degenerate gaps condition was weakened by showing that equilibration occurs provided that no energy gap is hugely degenerate.
This condition can be understood as
a way of preventing the situation where there is a subsystem which does not interact with the rest.

Let us finally point out that the equilibrium state introduced in Eq.~(\ref{eq:equilibrium-state}) is precisely the state that maximizes the von Neumann entropy given all the conserved quantities \cite{Gogolin2011}. This observation turns the \emph{principle of maximum entropy} into a consequence of the quantum dynamics. 
The principle of maximum entropy was introduced by Jaynes in Ref.~\cite{Jaynes1957}
and states that the probability distribution which best represents the current state of knowledge of the system is the one with largest entropy given the conserved quantities of the system.
We will come back in more detail to the Jaynes principle in the next section
when the thermalization for integrable systems is discussed.

\subsection{Thermalization. Emergence of Gibbs states in local Hamiltonians}
The next step in this program of justifying the methods of statistical mechanics from quantum mechanics 
is to tackle the issue of thermalization, i.~e.\ to understand why the equilibrium state is usually well described by a Gibbs state, which is totally independent of the initial state of the system, except for some macroscopic constraints such as its mean energy.
In Ref.~\cite{Riera2012}, a set of sufficient conditions for the emergence of Gibbs states is presented for the case of 
a subsystem $S$ that interacts weakly with its environment $B$ through a coupling $V$.
The Hamiltonian that describes such a situation is $H = H_S + H_B + V$.
These conditions are a natural translation of the three ingredients that enter the standard textbook proof of the canonical ensemble in classical statistical physics: 

\begin{enumerate}
\item  The \emph{equal a priory probability postulate} that has been replaced by typicality arguments in Section~\ref{sec:typicality}, and an equilibration postulate (such as the second law) that has been replaced by quantum dynamics in Section~\ref{sec:equilibration}. 

\item The assumption of \emph{weak-coupling}. Here, the standard condition from perturbation theory, $\norm{V}_\infty  \ll \textrm{gaps}(H)$, is not sufficient in  the thermodynamic limit, due to the fast growth of the density of states and the corresponding shrinking of the gaps in the system size. Instead, it is replaced with a physically relevant condition, $\norm{V}_\infty \ll k_B\ T$, which is robust in the thermodynamic limit.

\item An assumption about the \emph{density of states} of the bath
\footnote{The density of states of the bath $\varrho^B(E)$ is the number of eigenstates of the bath with energy close to $E$.}
, namely, that it grows faster than exponentially with the energy and that it can be locally approximated by an exponential.
\end{enumerate}

Note that the weak-coupling condition will not be satisfied in spatial dimensions higher than one for sufficiently large subsystems, since the interaction strength typically scales as the boundary of the subsystem $S$. 
This will be the case regardless of the strength of the coupling per particle or the relative size between $S$ and $B$. This should not be seen as a deficiency of the above results, but as a feature of strong interactions. 
Systems that strongly interact with their environment do not in general equilibrate towards a Gibbs state, in a similar way that the reduced state (of a part) of a Gibbs state need not have Gibbs form \cite{Ferraro2012,Kliesch2014}.
In this context, the findings of Ref.~\cite{Mueller2013} suggest that subsystems do not relax towards a local Gibbs state but to the reduction of the global Gibbs state; this is shown for translation-invariant quantum lattices with finite range but arbitrarily strong interactions.
The \emph{eigenstate thermalization hypothesis} (ETH) \cite{Srednicki1994,Rigol2008} gives further substance to this expectation.
ETH has several formulations. Its simplest one is maybe the one introduced in Ref.~\cite{Rigol2008}. It states that the expectation value $\bra{E_k}O\ket{E_k}$ of a few-body observable $O$ in an individual Hamiltonian eigenstate $\ket{E_k}$ equals the thermal average of $O$ at the mean energy $E_k$. 
Although ETH has been observed for some models, it is not true in general and it is well known to break down 
for integrable models (see \cite{Rigol2008} for an example with hard-core bosons and references in Ref.~\cite{Gogolin2015} for further examples).

In the same spirit, it has recently been proven that
a global microcanonical state (the completely mixed state of a energy shell subspace spanned by the Hamiltonian eigenstates with energy inside a narrow interval) and a global Gibbs state are locally indistinguishable for short range spin Hamiltonians off criticality, that is, when they have a finite correlation length \cite{Brandao2015b}. This represents a rigorous proof of the so called \emph{equivalence of ensembles}. If the Hamiltonian is not translationally invariant, the local indistinguishability between  canonical and microcanonical ensembles becomes a typical property of the subsystems, allowing for rare counterexamples.

Concerning the latter condition on the density of states of the bath, 
in Ref.~\cite{Keating2014} it has been
proven that the density of states of translational invariant spin chains 
 tends to a Gaussian in the thermodynamic limit, matching
the suited property of being well approximated by an exponential.
In Ref.~\cite{Brandao2015b}, the same statement is proven for any short ranged spin Hamiltonian.

Let us finally point out that not all systems thermalize.
For instance, \emph{integrable systems} are not well described by
the Gibbs ensemble. This is due to the existence of local integrals of motion, i.~e.\ conserved quantities, $Q_\alpha$ that keep the memory about the initial state. Instead, they turn to be described by the Generalized Gibbs Ensemble (GGE)
defined as
\begin{equation}
\tau_{GGE} \propto 
\exp\left(-\beta \left(H + \sum_\alpha\mu_\alpha Q_\alpha \right)\right)
\end{equation}
where the generalized chemical potential $\mu_\alpha$ is a Lagrange multiplier 
associated to the conserved quantity $Q_\alpha$ such that its expectation value is the same as 
the one of the initial state.
The GGE was introduced by Jaynes in Ref.~\cite{Jaynes1957} where he pointed out
that statistical physics can be seen as statistical inference
and an ensemble as the least biased estimate possible
on the given information. 
Nevertheless, note that any system has as many conserved quantities as the dimension of the Hilbert space, e.g. $Q_\alpha=\proj{E_\alpha}$.
If one includes all these conserved quantities into the GGE the ensemble obtained is the diagonal ensemble introduced in Eq.~(\ref{eq:equilibrium-state}).
Note that the description of the equilibrium state by the diagonal ensemble 
requires the specification of as many conserved quantities as the dimension of the Hilbert space, which scales exponentially in the system size, and becomes highly inefficient. 
A question arises here naturally, is it possible to provide an accurate description of the equilibrium state specifying only a polynomial number of conserved quantities? If so, what are these \emph{relevant conserved quantities} $Q_\alpha$ that allow for an accurate and efficient representation of the ensemble? This question is tackled in Ref.~\cite{Sels2014}.
There, it is argued that the relevant conserved quantities are the ones that make the GGE as close as possible to the diagonal ensemble in the relative entropy distance $D(\omega||\tau_{GGE})$, which in this particular case can be written as
\begin{equation}\label{eq:distance-diagonal-GGE}
D(\omega||\tau_{GGE})=S(\tau_{GGE})-S(\omega)\, ,
\end{equation}
where we have used that the diagonal ensemble and the GGE have by construction 
the same expectation values for the set of selected conserved quantities, i.~e.~$\Tr(Q_\alpha \tau_{GGE})=\Tr(Q_\alpha \omega)$.
Equation (\ref{eq:distance-diagonal-GGE}) tells us that the relevant conserved quantities are the ones the minimize the entropy $S(\tau_{GGE})$.
Note that in contrast to Jaynes approach, where entropy is maximized for a set of observables defined beforehand, here the notion of physically relevant is provided by how much an observable is able to reduce the entropy by being added into the set of observables that defines the GGE.

If instead of calculating the relative entropy between the diagonal ensemble and the GGE's we do it
with respect to the set of product states, i.~e.
\begin{equation}
T(\omega):=\min_{\pi_1, \pi_2,\ldots,\pi_n} D(\omega||\pi_1\otimes\pi_2\otimes\ldots\pi_n),
\end{equation}
then we obtain a measure of the \emph{total (multipartite) correlations} of the diagonal ensemble.
In Ref.~\cite{mbl} the scaling with system size of the total correlations of the diagonal ensemble has been shown to be
connected to ergodicity breaking and used to investigate the phenomenon of many-body localization. 

\subsection{Equilibration times}
Maybe the major challenge that is still open in the equilibration problem is to determine the equilibration timescale. It turns out that even if we
know that a system equilibrates, there are no relevant bounds on how long the equilibration
process takes. 
There could be quantum systems that are going
to equilibrate, but whose equilibration times are of the order of magnitude of the age of the universe, or alternatively, some systems, like glasses, which do not relax to equilibrium at all,
but have metastable states with long lifetimes. The problem of estimating equilibration timescales
is thus essential in order to have a full understanding of thermalization.

So far, progress on this issue has taken place from two different approaches. On the one hand, rigorous and completely general bounds on equilibration times have been presented in Ref.~\cite{Short2012}. Due to their generality, these bounds scale exponentially with the system size, leading to equilibration times of the age of the universe for macroscopic systems. On the other hand,
very short equilibration times have been proven for 
generic observables \cite{Malabarba2014}, Hamiltonians \cite{Goldstein2010a, Brandao2012,Vinayak,Cramer2012, Goldstein2013}, and initial states \cite{Hutter2013}.
In nature, systems seem to equilibrate in a time that is neither microscopic nor exponential in the system size.
A relevant open question is what properties of the Hamiltonians and operators  lead to reasonable equilibration time.
As a first step, in Ref.~\cite{Masanes2013}, a link between the complexity of the Hamiltonian's eigenvectors and equilibration time is presented. 
The result does not completely solve  the question, since the given bounds are not fulfilled by all Hamiltonians but only by a fraction of them,
and further research  in this direction is needed.

\subsection{Outlook}
The aim of this section has been to justify
that thermal states emerge in Nature for generic Hamiltonians.
To complete the picture presented here
we recommend the article \cite{Gogolin2015} 
where an extensive review of the literature on  foundations
of statistical mechanics is provided.

The main ideas presented here have also been widely studied in the
context of condensed matter physics, in which
systems are typically brought out of equilibrium by sudden (and slow) \emph{quantum quenches}:
the Hamiltonian of a system (that is initially in the ground state) is suddenly (or smoothly) changed in time.
We  recommend the review article \cite{Polkovnikov2011}
on non-equilibrium dynamics of closed interacting quantum systems.

Let us finish the section with
a list of some of the open problems that we consider most relevant in the field:
\begin{itemize}
\item \emph{Typicality for symmetric states}. Hamiltonians in nature are not generic but have symmetries.
Hence, the notion of typicality should be extended to physical states 
that are produced by symmetric Hamiltonians. 

\item \emph{Quantum notion of integrability}. One of the reasons why it is so difficult to extract strong statements on the equilibration and thermalization of many body quantum systems is the absence of a satisfactory quantum notion of
integrability \cite{Caux2011}. This leads first to some widespread confusion, since integrability is mentioned very often in the field of non-equilibrium dynamics, and second it does not allow us to classify quantum systems into classes with drastically different physical behaviour, like what occurs in classical mechanics.

\item \emph{Equilibration times}. Without bounds on the equilibration time scales, statements on equilibration become useless. As we have seen, the equilibration times are model dependent. 
We need then to understand how the equilibration times depend on the features of the Hamiltonian and the set of observables considered.

\item \emph{Relative thermalization.}
It was highlighted in Ref.~\cite{DelRio2014}
that local thermalization of a subsystem $S$, as  described here, is not enough to guarantee that $S$ will act as thermal bath towards another physical system $R$. In other words, imagine that we want to perform quantum thermodynamics on a reference system $R$, using $S$ as a thermal bath. 
To model a thermodynamic resource theory that recovers the laws of thermodynamics, it is not sufficient to demand that $S$ be in a local Gibbs state $\tau_S(\beta)$. Indeed we need $S$ to be thermalized relative to $R$, that is the the two systems should be uncorrelated, in global state, $\tau_S(\beta) \otimes \rho_R$. If this does not hold, then we cannot recover the usual thermodynamic monotones (for instance, there could be anomalous heat flows against the temperature gradient). Therefore, the relevant question for resource theories of thermodynamics is not only ``does $S$ thermalize locally after evolving together with an environment?'', but rather `does $S$ thermalize relative to $R$ after evolving together with an environment?', and the results discussed in this section should be generalized to that setting. First steps in this direction can be found in Ref.~\cite{DelRio2014}, where the authors use decoupling|a tool developed in quantum information theory|to find initial conditions on the entropies of the initial state that lead to relative thermalization.
\end{itemize}

\section{Resource theories}
\label{sec:resource_theories}
In the previous section we saw the progress that has been made in understanding how systems come to equilibrium, in particular thermal equilibrium, and as such a justification for the thermal state. In the rest of this review we will now take the thermal state as a given, and see what is the thermodynamics of quantum systems which start thermal or interact with thermal states. We will start from an operational point of view, treating the thermal state as a ``free resource'', a view inspired by other resource theories from quantum information. 

In this section we discuss the approach of thermodynamics as a resource theory in more detail.
Let us start by introducing the basic ideas behind resource theories that can be found in the literature, entanglement theory being the paradigmatic example. The first step is to fix the \emph{state space} $S$, which is usually compatible with a composition operation|for instance, quantum states together with the tensor product, in systems with fixed Hamiltonians.  The next step is to define the set of \emph{allowed state transformations}. For thermodynamics, these try to model adiabatic operations|like energy-preserving reversible operations, and contact with a heat bath. 

The set of allowed operations  induces a structure on the state space: we say that $\rho \to \sigma$ if there is an allowed transformation from $\rho$ to $\sigma$. The relation $\to$ is a  \emph{pre-order}, that is, a binary relation that is both reflexive ($\rho \to \sigma$) and transitive ($\rho \to \sigma$ and $\sigma \to \tau$ implies $\rho \to \tau$; this results from composing operations one after the other).

The task now is to find general properties of this structure. A paradigmatic example is looking for simple necessary and sufficient conditions for state transformations. The most general case are functions such that 
\begin{itemize}
\item $\rho \to \sigma \Rightarrow f(\rho,\sigma) \geq 0$ (that is, $ f(\rho,\sigma)\geq 0$ is a \emph{necessary} condition for state transformations), or
\item $f(\rho, \sigma) \geq 0 \Rightarrow \rho \to \sigma$ (that is, $ f(\rho,\sigma)\geq 0$ is a \emph{sufficient} condition for state transformations).
\end{itemize}
Often, we try to find necessary and sufficient conditions as functions that can be written like $f(\rho, \sigma) = g(\rho)- h(\sigma)$.
In the special case where $g=h$ for a necessary condition  ($\rho \to \sigma \Rightarrow g(\rho) \geq g(\sigma)$), we call $g$ a \emph{monotone} of the resource theory. For example, in classical, large-scale thermodynamics, the free energy is a monotone.

In order to quantify the cost of state transformations, we often fix a minimal unit in terms of a \emph{standard resource} that can be composed. For example, in entanglement theory the standard resource could be a pair of maximally entangled qubits, and in quantum thermodynamics we could take a single qubit (with a fixed Hamiltonian) in a pure state. The question then is `how many pure qubits do I need to append to $\rho$ in order to transform it into $\sigma$?' or, more generally, `what is the cost or gain, in terms of this standard resource, of the transformation $\rho \to \sigma$?' \cite{Lieb1999, Lieb2003, Faist2012}.

One may also try to identify special sets of states. The most immediate one would be the set of \emph{free states}: those that are always reachable, independently of the initial state. In standard  thermodynamics, these tend to be what we call equilibrium states, like Gibbs states. 
Another interesting set is that of \emph{catalysts}, states that can be repeatedly used to aid in transformations. We will revisit them shortly.

\subsection{Models for thermodynamics}
\label{sec:intro_models_thermodynamics}

Now that we have established the basic premise and structure of resource theories, we may look at different models for resource theories of thermodynamics, which vary mostly on the set of allowed operations.
In the good `spherical cow' tradition of physics, the trend has been to start from a very simple model that we can understand, and slowly expand it to reflect more realistic scenarios. In general there are two types of operations allowed: contact with a thermal bath and reversible operations that preserve some thermodynamic quantities. 
Each of those may come in different flavours. 

\subsubsection{Noisy and unital operations}
In the simplest case, all Hamiltonians are fully degenerate, so thermal states of any temperature are just fully mixed states, and there are no special conserved quantities. In this setting, thermodynamics inherits directly from the theory of noisy operations \cite{Horodecki2003}. We may model contact with a thermal bath as composition with any system in a fully mixed state, and reversible operations as any unitary operation. Furthermore, we assume that we can ignore, or trace out, any subsystem.
Summing up, noisy operations have the form 
\begin{eqnarray*}
\cT(\rho_A) = \tr_{A'} \left(U_{AB} \left[\rho_A \otimes \frac{\id_{B}}{|B|}\right] U_{AB}^\dagger\right),
\end{eqnarray*}
where $A'$ is any subsystem of $AB$ and $U$ is a unitary matrix.
Alternatively, we may allow only for maps that preserve the fully mixed state, $\cT_{A \to B}: \cT_{A \to B}(\frac{\id_A}{|A|}) = \frac{\id_B}{|B|}$, called \emph{unital maps} (an example would be applying one of two isometries and then forgetting which one).  The two sets|noisy operations and unital maps|induce the same pre-order structure in the state space.
In this setting, \emph{majorization} is a necessary and sufficient condition for state transformations \cite{Horodecki2003}.
Roughly speaking, majorization tells us which state is the most mixed. Let  ${\bf r} = (r_1, r_2, \dots , r_N)$ and ${\bf s} = (s_1, s_2, \dots , s_N)$ be the eigenvalues of two states $\rho$ and $\sigma$ respectively, in decreasing order. We say that ${\bf r}$ majorizes ${\bf s}$ if $\sum_{i=1}^k r_i \geq \sum_{i=1}^k s_i $, for any $k\leq N$.
In that case $\rho \to \sigma$;  monotones for this setting are called \emph{Schur monotone functions}, of which  information-theoretical entropy measures are examples~\cite{Dahlsten2011,DelRio2011,Faist2012, Brandao2013b, Gour2015}. For example, if $\rho$ majorizes $\sigma$, then the von Neumann entropy of $\rho$, $S(\rho)= - \tr(\rho \log_2 \rho)$, is smaller than $S(\sigma)$.
For a review, see \cite{Gour2015}.

\subsubsection{Thermal operations}
The next step in complexity is to let  systems have non-degenerate Hamiltonians. The conserved quantity is energy, and equilibrium states are Gibbs states of a fixed temperature $T$. For instance for a system $A$ with Hamiltonian $H_A$, the equilibrium state is $\tau_A(\beta) = e^{-\beta H_A} /\mathcal{Z}$. We can model contact with a heat bath as adding any system in a Gibbs state|this corresponds to the idealization of letting an ancilla equilibrate for a long time. 
A first approach to model physical reversible transformations is to  allow for unitary operations $U$ that preserve energy|either absolutely ($[U, H]=0$) or on average ($\tr  (H\rho)= \tr ( H\  (U \rho U^\dagger))$ for specific states).
Finally, we are again allowed to forget, or trace out, any subsystem. Together, these transformations are called thermal operations,
\begin{eqnarray*}
\cT(\rho_A) = \tr_{A'} \left(U_{AB} \left[\rho_A \otimes \tau_B(\beta) \right] U_{AB}^\dagger\right),
\end{eqnarray*}
where $A'$ is any subsystem of $AB$ and $U$ is an energy-conserving unitary \cite{Janzing2000}.
The monotones found so far are different versions of the free energy, depending on the exact regime  \cite{Brandao2011a,  Aberg2013a, Horodecki2013a,  Brandao2013b, Renes2014} (see Example~\ref{ex:free_energy_iid}). It is worth mentioning we can build necessary conditions for state transformations from these monotones, but sufficiency results are only known for classical states (states that are block-diagonal in the energy eigenbasis) \cite{Brandao2013b} and any state of a single qubit \cite{Lostaglio2014,PhysRevLett.115.210403}.  In the limit of a fully degenerate Hamiltonian, we recover the resource theory of noisy operations.

\begin{bigexample}{Free energy as a monotone.}{free_energy_iid}
This is an example of finding monotones for the resource theory of thermal operations \cite{Brandao2011a}. 
We are interested in finding the optimal rates of conversion between two states $\rho$ and $\sigma$, in the limit of many independent copies,
\begin{eqnarray*}
R(\rho\to\sigma) 
:= \sup_R \lim_{n\to \infty}
\rho^{\otimes n} \to \sigma^{\otimes R n}.
\end{eqnarray*}
If both $R(\rho\to\sigma), R(\sigma \to \rho) >0$, and these quantities represent  optimal conversion rates, then the process must be reversible, that is, $R(\rho\to\sigma) = 1/ R(\sigma\to\rho)$; otherwise we could build a perpetual motion engine, and the resource theory would be trivial. 
The idea is to use a minimal, scalable resource $\alpha$ as an intermediate step.
We can think of $\alpha$ as a currency: we will sell $n$ copies of $\rho$ for a number of coins, and use them to buy some copies of $\sigma$. 
To formalize this idea, we define the selling and buying cost of a state  $\rho$, or more precisely the distillation and formation rates, 
\begin{eqnarray*}
R^D(\rho) &:= R(\rho\to\alpha),\qquad
R^F(\rho) := R(\alpha\to\rho)= \frac1{R^D(\rho)}.
\end{eqnarray*}
In the optimal limit  we have the process
\begin{eqnarray*}
\rho^n \to \alpha^{n R^D(\rho)} \to \sigma^{n R^D(\rho) R^F(\sigma)}
\ \Leftrightarrow \ 
\rho^n \to \sigma^{n R(\rho\to\sigma)},
\end{eqnarray*}
which gives us the relation 
\begin{eqnarray*}
R(\rho\to\sigma) = \frac{R^D(\rho)}{R^D(\sigma)}.
\end{eqnarray*}
We have reduced the question to finding the distillation rate, which depends on the choice of $\alpha$.
For example, take $\rho$, $\sigma$ and $\alpha$ to be classical states (diagonal in the energy basis) of a  qubit with Hamiltonian $H=\Delta \pure 1$. For the currency, we choose $\alpha= \pure 1$.
The distillation rate is found by use of information-compression tools \cite{Brandao2011a}. It is given by the relative entropy between $\rho$ and the thermal state $\tau(\beta)$,
\begin{eqnarray*}
R^D(\rho) 
&=  D(\rho\|\tau(\beta))\\
&= \tr(\rho(\log \rho - \log \tau(\beta) ))\\
&= \beta (F_\beta(\rho) - F_\beta(\tau(\beta)) ),
\end{eqnarray*}  
where $F_\beta(\rho)= \avg E_\rho - \beta^{-1} S(\rho)$ is the free energy of $\rho$ at inverse temperature $\beta$.  
All in all, we find the conversion rate
\begin{eqnarray*}
R(\rho\to\sigma) = \frac{F_\beta(\rho) - F_\beta(\tau(\beta))}{F_\beta(\sigma) - F_\beta(\tau(\beta))}.
\end{eqnarray*}
Now we can apply this result to find a monotone for a single-shot scenario: in order to have $\rho\to\sigma$ we need in particular that $R(\rho\to\sigma) \geq 1$. In other words, we require $F_\beta(\rho) \geq F_\beta(\rho)$, thus recovering the free energy as a monotone for the resource theory of thermal operations. If we work directly in the single-shot regime, we recover a whole family of  monotones~\cite{Brandao2013b} based on quantum R\'enyi relative entropies~\cite{MuellerLennert2013}, of which the free energy is a member.
\end{bigexample}

\subsubsection{Gibbs-preserving maps} 
Following the example of the theory of noisy operations, we could try to replace these thermal operations with so-called Gibbs-preserving maps, that is, maps such that $\cT_{A\to B}(\tau_A(\beta)) =\tau_B(\beta)$.  This constraint is easier to tackle mathematically, and the two resource theories induce the same pre-order on classical states,  leading to a condition for state transformation called Gibbs-majorization (which is majorization after a rescaling of the eigenvalues) \cite{Horodecki2013a}.
However, Gibbs-preserving maps are less restrictive than thermal operations for general quantum states \cite{Faist2014b}. For example, suppose that you  have  a qubit with the Hamiltonian $H= E\ \pure1$, and you want to perform the transformation $\ket{1} \to \ket + = (\ket 0 +\ket1)/\sqrt{2}$. This is impossible through thermal operations, which cannot create coherence; yet there exists a Gibbs-preserving map that achieves the task. 
We may still use Gibbs-preserving maps to find lower bounds on performance, but at the moment we cannot rely on them for achievability results, as they are not operationally defined. 

\subsubsection{Coherence} The difference between thermal operations and Gibbs-preserving maps is not the only surprise that quantum coherence had in store for thermodynamics enthusiasts. The question of how to create coherence in the first place led to an intriguing discovery. In order to achieve the above transformation $\ket1\to \ket{+}$ through thermal operations, we need to draw coherence from a reservoir. A simple example of a coherence reservoir would be a doubly infinite harmonic oscillator, $H = \sum_{n=-\infty}^\infty n \Delta \ \pure n $, in a coherent state like $\ket \Psi = N^{-1}\sum_{n=a}^{a+N} \ket n$. Lasers approximate such reservoirs, which explains why we can use them to apply arbitrary transformations on quantum systems like ion traps. One may ask what happens to the reservoir after the transformation: how much coherence is used up? Can we use the same reservoir to perform a similar operation in a new system? The unexpected answer is that coherence is, in a sense, catalytic: while the state of the reservoir is affected, its ability to implement coherent operations is not \cite{Aberg2013}. What happens is that the state of the reservoir `spreads out' a little with each use, but the property that determines the efficacy of the reservoir to implement operations stays invariant. 
In more realistic models for coherence reservoirs, where the Hamiltonian of the reservoir has a ground state, the catalytic properties hold for some iterations, until the state spreads all the way down to the ground state. At that stage, the reservoir needs to be recharged with energy to pump up the state again. Crucially, we do not need to supply additional coherence. 
In the converse direction, we know that coherence reservoirs only are critical in the single-shot regime of small systems. Indeed, in the limit of processing many copies of a state simultaneously, the work yields of doing it with and without access to a coherence reservoir converge  \cite{Skrzypczyk2014}

\subsubsection{Catalysts}
The catalytic nature of coherence  raises more general questions about catalysts in thermodynamics. 
Imagine that we want to perform a transformation $\rho \to \sigma$ in a system $S$, and we have access to an arbitrary ancilla in any desired state $\gamma$. Now suppose that our constraint is that we should return the ancilla in a state that is $\epsilon$-close to $\gamma$:
\begin{eqnarray*}
 \rho_S \otimes \gamma_A \to \sigma_{SA}: 
 \qquad \|\sigma_A - \gamma_A \|_1 \leq \epsilon.
\end{eqnarray*}
The question is whether we can overcome the usual limits found in thermal operations by use of this catalyst. In other words, can we perform the above transformation in cases where $\rho\to \sigma$ would not be allowed? It turns out that if no other restrictions are imposed on the catalyst, then for any finite $\epsilon$ and any two states $\rho$ and $\sigma$, we can always find a (very large) catalyst that does the job \cite{Brandao2013b}. These catalysts are the thermodynamic equivalent of embezzling states in LOCC~\cite{Jonathan1999}. However, if we impose reasonable energy and dimension restrictions on the catalyst, we recover familiar monotones for state transformations 
\cite{Brandao2013b, Ng2014}. These restrictions and optimal catalysts result from adapting the concept of trumping relations on embezzling states \cite{Klimesh2007, Turgut2007}  to the thermodynamic setting. In particular, if we demand that $\epsilon \propto n^{-1}$, where $n$ is the number of qubits in the catalyst, we recover the free energy constraint for state transformations~\cite{Ng2014}.
A relevant open question, motivated by the findings of catalytic coherence, is what happens if we impose operational constraints on the final state of the catalyst. That is, instead of asking that it be returned $\epsilon$-close to $\gamma$, according to the trace distance, we may instead impose that its catalytic properties stay unaffected. It would be interesting to see if we recover similar conditions for allowed transformations under these constraints.

\subsubsection{Clocks}
All of resource theories mentioned allow for energy-preserving unitary operations to be applied for free. That is only the `first order' approach towards an accurate theory of thermodynamics, though. 
Actually, in order to implement a unitary operation, we need to apply a time-dependent Hamiltonian to the systems involved. To control that Hamiltonian, we require very precise time-keeping|in other words, precise clocks, and we should account for the work cost of using such clocks.
Furthermore, clocks are clearly out of equilibrium, and using them adds a source of free energy to our systems. Including them explicitly in a framework for work extraction forces us to account for changes in their state, and ensures that we do not cheat by degrading a clock and drawing its free energy. 
First steps in this direction can be found in \cite{Malabarba2014}.
There, the goal is to implement a unitary transformation in a system $S$, using a time-independent Hamiltonian. For this, the authors introduce an explicit clock system $C$ hat runs continuously, as well as a weight $W$ that acts as energy and coherence reservoir. The global system evolves under a time-independent Hamiltonian, designed such that the Hamiltonian applied on $S$ depends on the position of the clock|which effectively measures time. The authors show that such a construction allows us to approximately implement any unitary operation on $S$, while still obeying the first and second laws of thermodynamics. Furthermore, the clock and the weight are not degraded by the procedure (just like for catalytic coherence). 
In particular, this result supports the idea behind the framework of thermal operations: that energy-conserving unitaries can approximately be implemented for free (if we neglect the informational cost of designing the global Hamiltonian).
Note that this is still an idealized scenario, in which the clock is infinite-dimensional and moves like a relativistic particle (the Hamiltonian is proportional to the particle's momentum). A relevant open question is whether there exist realistic systems with the properties assigned to this clock, or alternatively how to adapt the protocol to the behaviour of known, realistic clocks. That direction of research  can be related to the resource theory of quantum reference frames \cite{Gour2008, Marvian2013,Frenzel2014, Lostaglio2014}.
An alternative direction would be to ask what happens if we do not have a clock at all|can we extract all the work from a quantum state if we are only allowed weak thermal contact? This question is studied (and answered in the negative, for general states) in Ref.~\cite{Wilming2014}.

\begin{bigexample}{Heat engines}{heat_engines}
The extreme case where one of our resources is in itself a second heat bath is of particular interest. This is a very natural scenario in traditional thermodynamics: steam engines used a furnace to heat a chamber, and exploit the temperature difference to the cooler environment. The study of this limit led to landmark findings like trains, fridges and general heat engines, and to theoretical results on the efficiency of such engines. One might wonder whether these findings can also be applied at the quantum scale, and especially to very small systems composed only of a couple of qubits~\cite{Ramsey1956, Scovil1959}. The answer is yes: not only  is  it possible to build two-qubit heat engines, but they achieve Carnot efficiency \cite{Linden2010, Brunner2011}. It is possible to build heat engines that do not require a precise control of interactions, in other words, that do not require a clock \cite{Linden2010, Popescu201d}.
\end{bigexample}

\subsubsection{Free states and passivity} 
It is now time to question the other assumption behind the framework of thermal operations: that Gibbs states come for free. There are two main arguments to support it: firstly, Gibbs states occur naturally under standard conditions, and therefore are easy to come by; secondly, they are useless on their own. The first point, typicality  of Gibbs states, is essentially the fundamental postulate of statistical mechanics: systems equilibrate to thermal states of Gibbs form. This assumption is discussed and ultimately justified from first principles in Section~\ref{sec:foundations}
The second point is more subtle. Pusz and Woronowicz first introduced the notion of passive states, now adapted to the following setting  \cite{Pusz1978, Lenard1978, Janzing2006}. Let $S$ be a system with a fixed Hamiltonian $H$, in initial state $\rho$. We ask whether there is a unitary $U$ that decreases the energy of $S$, that is
\begin{eqnarray*}
\tr (\rho H) > \tr (U \rho U^\dagger H). 
\end{eqnarray*}
If we can find such a unitary, then we could extract work  from $S$ by applying $U$ and storing the energy difference in a weight system. If there is no $U$ that achieves the condition above, then we cannot extract energy from $\rho$, and we say that the state is \emph{passive}.  The latter applies to classical states (i.e., diagonal in the energy basis) whose eigenvalues do not increase with energy. However, suppose that now we allow for an arbitrary number $n$ many copies of $\rho$ and  a global unitary $U_{\mathrm{gl}}$. The question becomes whether 
\begin{eqnarray*}
\tr (\rho H) > \frac 1n \tr (U_{\mathrm{gl}} \ \rho^{\otimes n} \ U_{\mathrm{gl}}^\dagger \ H_{\mathrm{gl}}), 
\end{eqnarray*}
where $H_{\mathrm{gl}}$ is the global Hamiltonian, which is the sum of the independent local Hamiltonians of every system.  If this is not possible for any $n$, we say that $\rho$ is \emph{completely passive}, and it turns out that only states of Gibbs form, $\rho= \tau(\beta)$ are completely passive. Moreover, Gibbs states are still completely passive if we allow each of the $n$ subsystems to have a different Hamiltonian, as long as all the states correspond to the same inverse temperature $\beta$.  This justifies the assumption that we may bring in any number and shape of subsystems in thermal states for free, because we could never extract work from them alone|another resource is necessary, precisely a state out of equilibrium. More formally, it was shown that if a resource theory allows only for energy-conserving unitaries and composition with some choice of free states, Gibbs states are the only choice  that does not trivialize the theory \cite{Halpern2014c, Brandao2013b}.

\subsubsection{Different baths}
The results outlined above suggest that thermodynamics can be treated as information processing under conservation laws, and so researchers began to experiment with other conserved quantities, like angular momentum \cite{Barnett2013, Vaccaro2011a, Weilenmann2015}, using the principle of maximum entropy to model thermal momentum baths.  The state of those baths has again an exponential Gibbs form, with operators like $L$ replacing $H$.  The same type of monotones emerged, and similar behaviour was found for more general conserved quantities  \cite{Halpern2014b, Halpern2014c}.

\subsubsection{Finite-size effects}
Another setting of practical interest is when we have access to a heat bath but may not draw arbitrary thermal subsystems from it. For instance, maybe we cannot create systems with a very large energy gap, or we can only thermalize a fixed number of qubits. In this case, the precision of state transformations is affected, as shown in~\cite{Reeb2014}, and we obtain effective measures of work cost that converge to the usual quantities in the limit of a large bath.  The opposite limit, in which all resources are large heat baths, leads to the idea of heat engines (Example \ref{ex:heat_engines}).

\subsubsection{Single-shot regime}
Some of the studies mentioned so far characterize the limit of many independent repetitions of physical experiments, and quantify things like the average work cost of transformations or conversion rates \cite{Brandao2011a, Skrzypczyk2014}. The monotones found (like the von Neumann entropy and the usual free energy) are familiar from traditional thermodynamics, because this regime approximates the behaviour of large uncorrelated systems.  
As we move towards a thermodynamic theory of individual quantum systems, it becomes increasingly relevant to work in the single-shot regime.  Some studies consider exact state transformations \cite{Lieb1999, Lieb2003, Weilenmann2015}, while others allow for a  small error tolerance \cite{Dahlsten2011,  DelRio2011, Egloff2012, Faist2012, Aberg2013a, Horodecki2013a,  Halpern2014b, Halpern2014c, Woods2015}. 
The monotones recovered correspond to operational entropy measures, like the smooth max-entropy (see Example~\ref{ex:erasure_intro}), and variations of a single-shot free energy that depend on the conservation laws of the setting; in general, they can be derived from quantum R{\'e}nyi relative entropies~\cite{MuellerLennert2013} between the initial state and an equilibrium state~\cite{Brandao2013b, Brandao2015}.
Single-shot results converge asymptotically to the traditional ones in the limit of many independent copies. The relation between single-shot and average regimes is studied via fluctuation theorems in \cite{Halpern2014}.

\subsubsection{Definitions of work}
In  classical thermodynamics, we can define work as some form of potential energy of an external device, which can be stored for later use. For instance, if  a thermodynamic process  results in the expansion of a gas against a piston, we can attach that piston to a weight,  that is lifted as the gas expands. We count the gain in gravitational potential energy as work|it is well-ordered energy that can later be converted into other forms, according to the needs of an agent.
A critical aspect is that at this scale fluctuations are negligible, compared to the average energy gain.
In the regime of small quantum systems, this no longer holds, and it is not straightforward to find a good definition of work.
Without a framework for resource theories of thermodynamics, a system for work storage is often left implicit. One option is to  assume that we can perform any joint unitary operation $U_{SB}$ in a system $S$ and a thermal bath $B$, and work is defined as the change in energy in the two systems manipulated,
$W:= \tr (H_{SB}\ \rho_{SB}) - \tr (H_{SB}\  U_{SB} \ \rho_{SB}\ U_{SB}^\dagger ) $,
where $H_{SB}$ is the (fixed) Hamiltonian of system and bath, and $\rho_{SB}$ the initial state~\cite{Reeb2014}.
Another example, inheriting more directly from classical thermodynamics, assumes that we can change the Hamiltonian of $S$ and bring it in contact with an implicit heat bath~\cite{AlickiHorodeckiHorodeckiHorodecki2004}; work at a time $t$ is then defined as 
$$W(t) := \int_0^t d t' \tr\left( \rho_S(t') \frac{d H_S(t')}{dt'} \right).$$
To study fluctuations around this average value, we consider work to be a random variable in the single-shot setting|this is explored in Section~\ref{sec:fluctuations}.
Note that in these examples work is not operationally motivated; rather it is defined as the change of energy that heat cannot account for. 
Resource theories of thermodynamics, with their conservation laws, force us to consider an explicit system $W$ for work storage. We act globally on $S\otimes W$, and we can define work in terms of properties of the reduced state of $W$.
One proposal for the quantum equivalent of a weight that can be lifted, for the resource theory of thermal operations, is a harmonic oscillator, with a regular Hamiltonian $H_W = \sum_n n \ \epsilon \ \pure n$. The energy gaps need to be sufficiently small to be compatible with the Hamiltonian of $S$; in the limit $\epsilon \to 0$ the Hamiltonian becomes $H_W = \int d x\ x \pure x $ \cite{Aberg2013a, Brunner2011}. Average work is defined as $\tr(H_W \  \rho_W^\mathrm{final}) - \tr(H_W \  \rho_W^\mathrm{initial})$, and fluctuations can be studied directly in the final state of the work storage system, $\rho_W$. This approach also allows us to observe other effects, such as the build up of coherences in $W$, and  of correlations between $W$ and $S$. 
Another advantage is that we can adapt the storage system to other resource theories: for instance, we can have an angular momentum reservoir composed of many spins, and count work in terms of polarization of the reservoir \cite{Vaccaro2011a}.
These approaches are critically analysed in Ref.~\cite{Gallego2015}; in particular, it is highlighted that they do not distinguish work from heat. For instance, thermalizing the work storage system may result in an increase of average energy, which is indiscriminately labelled as ``average work''.
In the same paper, an axiomatic approach to define work is  proposed, based on concepts from resource theories and interactive proofs. There, work is seen as a figure of merit: a real function assigned to state transformations, $\mathcal W(\rho \to \sigma)$.
Starting from a couple of assumptions, the authors derive properties of acceptable work functions $\mathcal W$: for instance, that they can be written as the difference between a monotone for initial and final state, $\mathcal W(\rho \to \sigma) = g(\rho) - g(\sigma)$. The free energy is an example of such a valid work function.

\begin{bigexample}{Landauer's principle}{erasure_intro}
 How much energy is needed to perform logical operations? What are the ultimate limits for heat dissipation of computers? These questions lie at the interface between thermodynamics and information theory, are of both foundational and practical interest. As Bennett realized, all computations can be decomposed into reversible operations followed by the erasure of a subsystem~\cite{Bennett1973}. If we assume that the physical support of our computer is degenerate in energy, we recover the setting of noisy operations, in which unitaries are applied for free. That way, the thermodynamic cost of computation is simply the cost of erasure, which is defined as taking a system from its initial state $\rho$ to a standard, predefined pure state $\ket0$ (like when we format a hard drive). Rolf Landauer first proposed that the work cost of erasing a completely unknown bit of information (think of a fully mixed qubit) in an environment of temperature $T$ is $k_B T \ln 2$ \cite{Landauer1961}. That very same limit was also found for quantum systems, in the setting of thermal operations \cite{Skrzypczyk2013, Reeb2014}, for the ideal case of an infinitely large heat bath and many operations; finite-size effects are analysed in Ref.~\cite{Reeb2014}.  
 
Using Landauer's principle as a building block, we can approach the more general question of erasing of a system that is not in a completely unknown state, but rather about which we have partial information. For example, imagine that we want to perform an algorithm in our quantum computer, and then erase a subsystem $S$ (which could be a register or ancilla). The rest of our computer may be correlated with $S$, and therefore we can use it as a memory $M$, and use those correlations to optimize the erasure of $S$. In short, we want to take the initial state $\rho_{SM}$ to $\pure0_S \otimes \rho_M$, erasing $S$ but not disturbing $M$. It was shown~\cite{DelRio2011,Faist2012} that the optimal work cost of that transformation is approximately $\hmax^\epsilon(S|M)_\rho k_B T \ln 2$,  where $\epsilon$ parametrizes our error tolerance and $\hmax^\epsilon(S|M)_\rho$ is the smooth max entropy,  a conditional entropy  measure that measures our uncertainty about the state of $S$, given access to the memory $M$. It converges to the von Neumann entropy in the limit of many independent copies. In the special case where $S$ and $M$ are entangled, it may become negative|meaning that we may gain work in erasure, at the cost of correlations. Not incidentally, these results use quantum information processing techniques to compress the correlations between $S$ and $M$ before erasure; after all, `information is physical' \cite{Landauer1991}.
\end{bigexample}

\subsection{Generalizing resource theories}
\label{sec:intro_generalizing}

Let us now abstract from particular resource theories, and think about their common features, and how we may generalize them.

\subsubsection{Starting from the pre-order} 
As mentioned before, the set of allowed transformations imposes a pre-order structure $(S, \leq)$ on  the state space $S$.  One direction towards exploring the concept of resource theories could be to start precisely from such a pre-order structure. That was the approach of Carath\'eodory, then Giles and later Lieb and Yngvason, who pioneered the idea of resource theories for thermodynamics~\cite{Giles1964, Lieb1999, Lieb2003,cata1909}. 
In their work, the set of allowed transformations is implicitly assumed, but we work directly with an abstract state space equipped with a preorder relation. They were largely inspired by classical, macroscopic thermodynamics, as one may infer from the conditions imposed on the state space,
but their results can be applied to thermodynamics of small quantum systems
\cite{Weilenmann2015}. 
Assuming that there exist minimal resources that can be scaled arbitrarily and act as `currency', the authors obtain monotones for exact, single-shot state transformations. When applied to the pre-order relation on classical states that emerges from thermal operations, these monotones become single-shot versions of the free energy~\cite{Weilenmann2015}.

\subsubsection{Starting from the set of free resources}
In~Ref.~\cite{Brandao2015} general quantum resource theories are characterized based on the set of free resources of each theory. Assuming that the set of free states is well-behaved (for instance, that it is convex, and that the composition of two free states is still a free state), they show that the relative entropy  between a resource and the set of free states is a monotone. This is because the relative entropy is contractive (non-increasing under quantum operations); the same result applies to any contractive metric. Finally, they find an expression for the asymptotic value of a resource  in terms of this monotone: the conversion rate between two resources is given by the ratio between their asymptotic value.

\subsubsection{In category theory} Ref.~\cite{Coecke2014}, and more recently Ref.~\cite{Fritz2015} have generalized the framework of resource theories to objects known as symmetric monoidal categories. These can represent essentially any resource that can be composed (in the sense of combining copies of different resources, like tensoring states in quantum theory). The authors consider both physical states and processes as possible resources. After obtaining the pre-order structure from a set of allowed operations, resource theories can be classified according to several parameters. For instance, the authors identify quantitative theories (where having more of a resource  helps, like for thermal operations) and qualitative ones (where it helps to have many different resources). They find expressions for asymptotic conversion rates in different regimes and, crucially, give varied examples of resource theories, within and beyond quantum theory, showing just how general this concept is.

\subsubsection{Resource theories of knowledge} 
In Ref.~\cite{DelRio2015}, emphasis is given to the subjective knowledge of an observer. The framework introduced there allows us to embed macroscopic descriptions of reality into microscopic ones, which in turn lets us switch between different agents' perspectives, and see how traditional large-scale thermodynamics can emerge from  quantum resource theories like thermal operations. It also allows us to combine and relate different resource theories (like thermodynamics and LOCC), and to infer the structure of the state space (like the existence of subsystems or correlations) from modularity and commutativity of transformations.

\subsection{Outlook}

In the previous sections we identified several open problems. These can be grouped into two main directions:
\begin{itemize}
\item \emph{Quantumness: coherence, catalysis and clocks.} It remains to find optimal coherent catalysts and clocks under realistic constraints (a generalization of Ref.~\cite{Ng2014}). This would give us a better understanding of the thermodynamic power and limitations of coherent quantum states. It would also allow us to account for all costs involved in state transformations. 

\item \emph{Identifying realistic conditions.}  We have been very good at defining sets of allowed transformations that are analogous to those of traditional thermodynamics, and recover the same monotones (like the free energy) in the limit of large, uncorrelated systems. The original spirit of thermodynamics, however, was to find transformations that were easy and cheap to implement for experimenters|for instance, those whose cost did not scale with the relevant parameters.  
In order to find meaningful resource theories for individual quantum systems, it is again imperative to turn to concrete experimental settings and try to identify easy and cheap transformations and resources. At this stage, it is not yet clear whether these will correspond to thermal operations, time-independent Hamiltonians, or another model of quantum thermodynamics|in fact it is possible that they vary depending on the experimental realization, from superconducting qubits to ion traps.

\end{itemize}


\section{Entanglement theory in thermodynamic settings}
\label{sec:entanglement}
In the previous sections we have established how quantum information can be used to understand the very foundation of thermodynamics, from the emergence of thermal states to the resource theory of manipulating these with energy conserving unitaries. We have seen that phrasing thermodynamics as a resource theory can elucidate the meaning of thermodynamic quantities at the quantum scale, and how techniques originally developed for a resource theory of communication can facilitate this endeavour. The motivation behind this approach is a very practical one: finding the ultimate limitations of achievable transformations under restrictions that follow from the nature of the investigated system that naturally limits the set of operations we can perform. As quantum information processing is becoming increasingly applied, we also need to think about fundamental restrictions to quantum information itself, emerging from unavoidable thermodynamic considerations. There has thus been an increased interest in investigating scenarios of quantum information processing where thermodynamic considerations cannot be ignored. From fundamental limitations to the creation of QIP resources to their inherent work cost. In this section we try to give a brief overview over recent developments in this intersection with a focus on the paradigmatic resource of QIP: entanglement.


\subsection{Correlations and entanglement under entropic restrictions}

Entanglement theory is in itself one of the most prominent examples of resource theories. Entanglement, a resource behind almost all tasks in quantum information processing, is hard to create and once distributed can only decrease. Thus in entanglement theory classically correlated states come for free and local operations are considered cheap, which singles out entanglement as the resource to overcome such limitations. These limitations and resources are of course very different to the resources and tasks explored in the previous sections. A comprehensive comparison between the principles behind these and more general resource theories is made in Ref.~\cite{HorodeckiOppenheim2013} and as examples of a more abstract treatment in Ref.~\cite{Coecke2014}.

Such resource theories are always designed to reflect specific physical settings, such as local operations and classical communication (LOCC) \cite{Bennett1996b} as a natural constraint for communication. It is therefore unavoidable that when describing various physical circumstances these resource theories can be combined yielding hybrid theories. One natural example is the desire to process quantum information in a thermodynamic background. Ignoring limitations coming from available energies in a first step this leads to the task of producing resources for computation (such as entanglement or correlation) at a given entropy. Some of the first considerations in this direction were motivated by the prospect of using nuclear magnetic resonance (NMR) for quantum computation. Due to non-zero temperature, i.e. non-trivial restrictions on the entropy of the state, such systems would always be fairly close to the maximally mixed state.

In  this context the most natural question to ask, is whether a unitary transformation is capable of entangling a given input state. As a precursor to studying the possibility of entangling multipartite states, the complete solution for two qubits was found in Ref.~\cite{MEMS1} and later decent bounds on bipartite systems of arbitrary dimension were presented in Ref.~\cite{MEMS2}.

Another pathway was pursued by Refs.~\cite{sepball1,sepball2,sepball3}, where with NMR quantum computation in mind, volumes of separable states around the maximally mixed state were identified. These volumes imply that if any initial state is in close proximity of the maximally mixed state, there can be no chance of ever creating entanglement in such states, as the distance from the maximally mixed state is invariant under unitary transformations. Further improvements in terms of limiting temperatures were obtained in Ref.~\cite{Chuang2005}.

The question of whether a given state can be entangled under certain entropy restrictions clearly relies only on the eigenvalue spectrum of the considered state, as the best conceivable operation creating entanglement is a unitary one (which leaves eigenvalues unchanged). These questions were further pursued under the name of ``separability from spectrum´´ in Refs.~\cite{EFS1,EFS2,EFS3}. One of the main results important in the context of quantum thermodynamics is the following:
A state with eigenvalues $\lambda_i$, ordered by size, i.e. $\{\lambda_i\geq\lambda_{i+1}\}$ can be entangled by an appropriate unitary if
\begin{eqnarray}
\lambda_1>\lambda_{d-1}+2\sqrt{\lambda_{d-2}\lambda_{d}}\,.
\end{eqnarray}
More importantly, for $2\times m$ dimensional states, this condition is not only sufficient, but also necessary \cite{EFS3}.

\begin{boxfigure}{Creating correlations between local thermal states}{entangling_operations}

 \includegraphics[width= \textwidth]{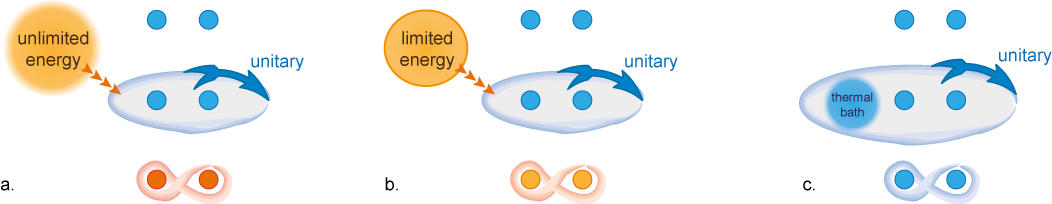}

\tcblower
Two systems, initially in local, product thermal states of temperature $T$, are correlated  using a) any unitary \cite{Jevtic2012b}, b) a unitary changing the average energy by at most $\Delta E$ \cite{Huber2014b} or c) a unitary on the systems and another bath at the same temperature $T$ \cite{Bruschi2014}. The local states of the two systems may heat up in the process: $\Delta T$ is infinite in a), a function of $\Delta E$ in b) and forced to be zero in c). In general, for setting b),  the optimal conversion rate of average energy into mutual information is still unknown. 
\end{boxfigure}

Moving beyond the mere presence of entanglement in the unitary orbit of input states, one encounters an intrinsic difficulty of properly quantifying the entanglement created. There is a whole ``zoo´´ of entanglement measures \cite{horodeckiqe,Siewert2014} and only in the bipartite case there is a unique "currency" known, i.e. a paradigmatic resource state from which all other states can be created via LOCC (although recent progress has been made in the four qubit case, where it has been shown that after exclusion of a measure zero set, such a set of resource states can indeed be identified \cite{Vicente2013}).

In any case one can at least study general correlations with a clear operational interpretation, such as the mutual information, which has been performed in Refs.~\cite{Jevtic2012,Jevtic2012b,Funo2013}. In these papers the authors have, among other things, identified minimally and maximally correlated states in the unitary orbit of bipartite systems. It turns out that at least here the entropy poses only a rather trivial restriction and for any $d$-dimensional state $\rho$ a mutual information of $I_{\rho}(A:B)=2\log_2(d)-S(\rho)$ can be achieved via global unitary rotations.

Exploiting these results Ref.~\cite{Huber2014b} continued to study the generation of correlations and entanglement under entropic restrictions for multipartite systems. Inspired by the idea of thermal states as a free resource, the authors consider a multipartite system initially in a thermal state. They ask what is the highest temperature $T_{ent}$ at which entanglement can still be created, it scales with the dimension of the partitions and quantify the inherent cost in terms average energy change (see example (\ref{ex:entanglement_2qubit}) for an exemplary two qubit energy cost). By introducing concrete protocols, i.e. unitary operations, the authors show that bipartite entanglement generation across all partitions of $n$-qubits is possible iff $k_BT/E<n/(2\ln(1+\sqrt{2}))$ and genuine multipartite entanglement across all parties can be created if $k_BT/E<n/(2\ln(n))+\mathcal{O}(n/\ln(n)^2)$.

\subsection{Correlations and entanglement in a thermodynamic background}

In the context of thermodynamics the previous subsection can be viewed as a very special case of operating on closed systems with an unlimited external energy supply or a fully degenerate Hamiltonian. As elaborated in Section \ref{sec:resource_theories} of the review this does not encompass the whole potential of thermodynamic transformations. If the necessary correlating unitary does not conserve the total energy, we should account for the difference in average energy between initial and final states. Taking into account also the average energy cost reveals an intrinsic work value of correlations and entanglement in general. This fundamental fact was first quantified in Ref.~\cite{Huber2014b}).

\begin{bigexample}{Entangling two qubits}{entanglement_2qubit}
Creating entanglement from thermal states will always cost some energy. For the simplest case of entangling two qubits with energy gap $E$ at zero temperature one can find a closed expression, e.g. for the concurrence, in terms of the invested average energy $\Delta E=W$:
\begin{eqnarray}
C(W)=\sqrt{\frac{W}{E}\left(2-\frac{W}{E}\right)}\nonumber\,,
\end{eqnarray}
\end{bigexample}

Accounting for the average energy change in the unitary orbit of initial quantum states however still does not encompass the whole potential of thermodynamic resource theories. Thermal operations on target states can also make use of a thermal bath at temperature $T$ and thus can also reduce the entropy of the target system. Disregarding energy costs in this context of course yields the rather trivial result that any quantum information processing resource can be produced, simply by cooling the system (close) to the ground state and then performing the adequate global unitary operation on it. Taking into account the free energy costs of correlating transformations, Ref.~\cite{Bruschi2014} has shown that every bit of correlation embodies an intrinsic work value proportional to the temperature of the system. For mutual information this yields a relation akin to Landauer's principle for the work cost of creating correlations $W_{cor}$,
\begin{eqnarray}
W_{cor}\geq k_B T I_\rho (A:B)\,,
\end{eqnarray}
and it implies a general free energy cost of entanglement that is bounded from above and below for the bipartite case in Ref.~\cite{Bruschi2014}. All previous considerations are illustrated in Fig.~\ref{fig:entangling_operations}.

That extractable work can be stored in correlations is by no means a purely quantum phenomenon. Even classical correlations can store work in situations where local work extraction is impossible. In Ref.~\cite{Perarnau-Llobet2014a} the quantum vs. classical capacity for storing extractable work purely in correlations was compared. For two qubits twice as much work can be stored in entangled correlations as the best possible separable (or even classical, which turns out to be the same) correlations admit (a fact that is also mentioned in Ref.~\cite{DelRio2014} in a different setting). However the difference between separably encoded work from correlations $W_{sep}$ and the maximally possible work in correlations $W_{max}$ scales as
\begin{eqnarray}
\frac{W_{sep}}{W_{max}}=1-\mathcal{O}(n^{-1})\,,
\end{eqnarray}
i.e. the quantum advantage vanishes in the thermodynamic limit of large systems.\\
Concerning the extractable work from correlations one can also find seemingly contrary results if the figure of merit changes. The above considerations apply only if the target is an extraction of average energy or standard free energy, partially neglecting the details of the work distribution arising in the receiving system (detailed considerations of such work distribution fluctuations will be discussed in section (\ref{sec:fluctuations})). One can just as well be interested in a guaranteed amount of work. If that is the case one can arrive at more restrictions concerning work extraction as also recently demonstrated in ~\cite{Woods2015}. Curiously in Ref.~\cite{mueller2014} it was shown, however, that these restrictions can be overcome by considering $k$ initially uncorrelated catalysts that build up correlations in the process. In that context one can extract more deterministic work and can thus regard the stochastic independence of the input catalysts as a resource for work extraction, which is quite contrary to the case considered before and the thermodynamic limit.

A different, but very related, setting exploring work gain from correlations is studied in the context of quantum feedback control. Here the task is rather to quantify the inevitable work cost arising from information gain in the process of a measurement. As in order to measure a system one needs to correlate with the system in question it follows intuitively that this scenario will also always induce work cost related to bipartite correlations between the system and the memory storing the information gain about the system. Here the work cost coming purely from correlations was quantified in Ref.~\cite{Funo2013}, building upon older results on the inevitable work cost of quantum measurements \cite{Groenewold1971,Ozawa1986,Sagawa2008,Sagawa2009} and Landauer's principle. To model the necessary feedback control, the authors included a general model of a quantum memory upon which projective measurements can be performed. The authors also studied the possible work gain from bipartite quantum states in this context. Denoting the state of the memory as $\rho_M$ the authors find an upper bound on the work gain (defined as the work extracted from both subsystems minus the work cost of the measurements and subsequent erasure of the quantum memory) as
\begin{align}
W_{net}\leq k_B T(I_\rho(A:B)-I_\rho(A:B|\rho_M))-\Delta F_\beta(\rho)\,.
\end{align}

\subsection{Thermodynamics under locality restrictions}

In the previous subsections we have reviewed the prospect of creating quantum information processing resources in a thermodynamic background. The other obvious connection between the resource theories of entanglement and thermodynamics is taking the converse approach. Here one is interested in thermodynamic operations under additional locality restrictions.

In Ref.~\cite{Oppenheim02} the difference between the extractable work from bipartite quantum states in thermodynamics both with and without locality restrictions was studied. The resulting difference, called the \emph{work deficit}, can be bounded via
\begin{eqnarray}
\Delta =\max[S(\rho_A),S(\rho_B)]-S(\rho_{AB})\,,
\end{eqnarray}
which for pure states coincides with entanglement of formation (or any other sensible choice of entanglement measure that all reduce to the marginal entropy in case of pure states). In the above equation it is assumed that bits which are sent down the communication channel are treated as classical in the sense that they are only dephased once, and not again in a second basis. This interplay led to subsequent investigations into the thermodynamic nature of entanglement in Ref.~\cite{Oppenheim02}, where analogies between irreversible operations in thermodynamics and bound entanglement were drawn, and to concrete physical scenarios satisfying this bound in Ref.~\cite{AlickiHorodeckiHorodeckiHorodecki2004}.

\subsection{Entanglement resources in thermodynamic tasks}

Apart from resource theory inspired questions, one might study the role of informational quantities through their inevitable appearance in thermodynamic operations at the quantum level. For instance the role of entangling operations and entanglement generation in extracting work from multiple copies of passive states , i.e. states where no local work extraction is possible \cite{Pusz1978,Lenard1978}, has attracted some attention recently. The implied fact that global unitary operations are required to extract work indicates some form of non-local resource being involved in the process.

In general passive states are always diagonal in the energy eigenbasis \cite{Pusz1978,Lenard1978}, which implies that one starts and ends the protocols with diagonal states. In these scenarios the individual batteries from which work is to be extracted are considered non-interacting, directly implying the separability of initial and final states in these protocols. Nonetheless the fact that local unitaries can never extract any work from copies of passive states directly implies that entangling unitaries enable work extraction from such states \cite{AlickiFannes13}. In that sense entangling power of unitaries can be seen as a resource for work extraction purposes (which in conventional thermodynamic resource is of course considered a free operation).

In Ref.~\cite{HPMA13} the role of quantum resources in this context was further explored. While it is true that the ability to perform entangling unitaries is required for this particular work extraction problem, this does not imply that any entanglement is ever generated in the process. In fact the whole procedure can dynamically be implemented without ever generating the slightest bit of entanglement \cite{HPMA13}, however the most direct transformation can considerably entangle the systems in the process. In Ref.~\cite{Binder2015} it was demonstrated that if the work per unit time (power) is considered with cyclic operations in mind then a quantum advantage for charging power can be achieved.

\subsection{Using thermodynamics to reveal quantumness}

That entanglement plays a special role in quantum many-body physics is a well established fact that has received adequate attention in numerous publications (see e.g. Ref.~\cite{Lewenstein2007} and the extensive list of citations therein). In this topical review we want to at least mention a related question that connects quantum thermodynamics directly with entanglement theory: The possibility to use thermodynamic observables to reveal an underlying entanglement present in the system. At zero temperature it is already known that many natural interaction Hamiltonians have entangled ground states (in fact often many low energy eigenstates even of local Hamiltonians feature entanglement). This fact can be exploited to directly use the energy of a system as an entanglement witness, even at non-zero temperatures \cite{Vedral2004}. Intuitively this can be understood through the fact that a low average energy directly implies that the density matrix is close to the entangled ground state. If this distance is sufficiently small that can directly imply entanglement of the density matrix itself. The known results and open questions of this interplay including Refs.~\cite{Dowling2004,Brukner2004a,Toth2005,Guhne2005,Guhne2006,Anders06,Vedral2009a,Wiesniak2008} are also discussed in the review Ref.~\cite{Amico2008a}. Furthermore, other macroscopic thermodynamic quantities can also serve as entanglement witnesses through a similar intuition, such as e.g. the magnetic susceptibility \cite{Wiesnak2005} or the entropy \cite{Bauml2015b}.

\subsection{Outlook}

 Resource theories always have their foundation in what we believe to be hard/impossible to implement and what resources allow us to overcome such limitations. As such they always only capture one specific aspect of the physical systems under investigation. The results outlined in this section emphasise the fact that thermodynamic constraints have drastic consequences for processing quantum information and that locality constraints will change thermodynamic considerations at the quantum scale. One path to explore could now be a consistent resource theory that adaptively quantifies possible resources from different restrictions. This would furthermore elucidate the exact role played by genuine quantum effects, such as entanglement, in thermodynamics.
\section{Quantum Fluctuation relations and quantum information}
\label{sec:fluctuations}
\subsection{Introduction}

The phenomenological theory of thermodynamics successfully describes the equilibrium properties of macroscopic systems ranging from refrigerators to black holes that is the domain of the large and many. By extrapolating backwards, from the domain of the `many' to the `few', we venture further from equilibrium into a regime where both thermal and quantum fluctuations begin to dominate and correlations proliferate. One may then ask the question - what is an appropriate way to describe this blurry world which is dominated by deviations from the average behaviour?

One way to describe the thermodynamics of small systems where fluctuations cannot be ignored is by using the framework of stochastic thermodynamics~\cite{Sekimoto2010}. In this approach the basic objects of traditional statistical mechanics such as work and heat are treated as stochastic random variables and hence characterised by probability distributions. Over the last 20 years various approaches have lead to sets of theorems and laws, beyond the linear response regime, which have revitalised the already mature study of non-equilibrium statistical mechanics. Central to these efforts are the fluctuation relations that connect the non-equilibrium response of a system to its equilibrium properties. A wealth of results have been uncovered in both the classical and the quantum regimes and the interested reader is directed to the excellent reviews on the topics~\cite{Seifert2012,Esposito2009,Campisi2011}. Here we focus on  aspects of this approach that have been specifically influenced by concepts in quantum information, or show promise for symbiosis. We hope that by reviewing the existing contributions as well as suggesting possible research avenues, further cross fertilisation of the fields will occur. 

To begin with, it is useful to illustrate how the probability distributions of a thermodynamic variable like work is defined. Consider a quantum system with a time-dependent Hamiltonian $H(\lambda(t))$, parametrized by  the externally controlled \emph{work parameter} $\lambda(t)$. The system is prepared in a thermal state by allowing it to equilibrate with a heat bath at inverse temperature $\beta$ for a fixed value of the work parameter $\lambda(t< t_\textrm{i})=\lambda_{\textrm{i}}$. The initial state of the system is therefore the Gibbs state,
$$\tau(\lambda,\beta):=
 \frac{e^{-\beta H(\lambda_i)}}{\mathcal{Z}_\beta(\lambda_\textrm{i})}$$
 At $t=t_\textrm{i}$  the system-reservoir coupling is removed and a fixed, reversible protocol is performed on the system taking the work parameter from its initial value $\lambda_\textrm{i}$ to the final value $\lambda_\textrm{f}$ at a later time  $t=t_\textrm{f}$. The initial and final Hamiltonians are defined by their spectral decompositions
 \begin{eqnarray*}
 H(\lambda_\textrm{i})&=\sum_n E_n(\lambda_\textrm{i})\ \proj{\psi_n}\\
 \mathrm{and}\quad 
 H(\lambda_\textrm{f})&=\sum_m E_m(\lambda_\textrm{f})\ \proj{\phi_m}
 \end{eqnarray*}
 respectively, where $\ket{\psi_n}$ ($\ket{\phi_m}$) is the $n$th ($m$th) eigenstate of the initial (final) Hamiltonian with  eigenvalue $E_n(\lambda_\textrm{i})$,$E_m(\lambda_\textrm{f})$. The  protocol connecting the initial and final Hamiltonians generates the unitary evolution operator $U(t_\textrm{f},t_\textrm{i})$, which in general has the form
\begin{eqnarray}
U(t,t_\textrm{i})=\mathbf{T}_\rightarrow \exp\left[-i\integral{t'}{t}{t_\textrm{i}} H(\lambda(t'))\right],
\end{eqnarray}
where $\mathbf{T}_\rightarrow$ denotes the time-ordering operation. We stress here that, in this framework, one typically assumes that the system is initially prepared in a thermal state but after the unitary protocol the system is generally in a non-equilibrium state. 

The work performed (or extracted) on (or from) the system as a consequence of the protocol is defined by the outcomes of two projective energy measurements~\cite{Talkner2007}. The first, at $t=t_\textrm{i}$, projects onto the eigenbasis of the initial Hamiltonian $H(\lambda_\textrm{i}$), with the system in thermal equilibrium. The system then evolves under the unitary operator $U(t_\textrm{f},t_\textrm{i})$  before a second projective measurement is made onto the eigenbasis of the final Hamiltonian $H(\lambda_\textrm{f})$ at $t=t_\textrm{f}$. The joint probability of obtaining the outcome $E_n(\lambda_\textrm{i})$ for the initial measurement followed by  $E_m(\lambda_\textrm{f})$ for the final one is easily shown to be
\begin{align}
p(n,m)={}&\frac{e^{-\beta E_n(\lambda_\textrm{i})}}{\mathcal{Z}(\lambda_\textrm{i})}  \ \vert \bra{\phi_m} U(t_\textrm{f},t_\textrm{i}) \ket{\psi_n} \vert^2.
\label{eq:condprob}
\end{align}
Accordingly, the quantum work distribution is defined as
\begin{align}
P_\textrm{F}(W)=\sum_{n,m} p(n,m) \ \delta\left(W-\left[\, E_m(\lambda_\textrm{f})-E_n(\lambda_\textrm{i}) \, \right] \, \right).
\label{eq:qworkdist}
\end{align}
where $\delta$ is the Dirac delta function. For reasons which will become clear shortly we use the subscript $F$ to denote `forward' protocol. Physically, Eq.~\ref{eq:qworkdist} states that the work distribution consists of the discrete number of allowed values for the work $\left( E_m(\lambda_\textrm{f})- E_n(\lambda_\textrm{i}) \right)$  weighted by the probability $p(n,m)$ of measuring that value in a given realisation of the experiment. The quantum work distribution therefore encodes fluctuations in the measured work arising from thermal statistics (first measurement) and from quantum measurement statistics  (second measurement).

In order to understand what is meant by a fluctuation theorem, we introduce a backward process which is the time reversed protocol of the forward one previously defined. 
Now $P_\textrm{B}(W)$ is the work distribution corresponding to the \emph{backward process}, in which the system is prepared in the Gibbs state of the final Hamiltonian $H(\lambda_\textrm{f})$ at $t=0$ and subjected to the time-reversed protocol that generates the evolution $\Theta U(t_\textrm{f},t_\textrm{i}) \Theta^\dagger$, where $\Theta$ is the anti-unitary time-reversal operator. It turns out that the following theorem holds, the Tasaki-Crooks relation \cite{Crooks1999,Tasaki2000}, 
\begin{eqnarray}
\frac{P_\textrm{F}(W)}{P_\textrm{B}(-W)}=e^{\beta(W-\Delta F)},
\label{eq:crooks}
\end{eqnarray}
which shows that, for any closed quantum system undergoing an arbitrary non-equilibrium transformation, the fluctuations in work are related to the equilibrium free energy difference for the corresponding isothermal process between the equilibrium states $\tau(\lambda_i)$ and
$\tau(\lambda_f)$,
\begin{eqnarray}
\Delta F= \frac 1 \beta  \  \ln 
 \left (
\frac {\mathcal{Z}_\beta(\lambda_\textrm{i})}
   {\mathcal{Z}_\beta(\lambda_\textrm{f})}
\right).
\end{eqnarray}

This relationship is further emphasized by a corollary to Eq.~\ref{eq:crooks} known as the Jarzynski equality \cite{Jarzynski1997},
\begin{eqnarray}
\int \textrm{d}W P_\textrm{F}(W)e^{-\beta W}=\langle e^{-\beta W} \rangle=e^{-\beta \Delta F}
\end{eqnarray}
which states that $\Delta F$ (of the corresponding isothermal process) can be extracted from by measuring the exponentiated work. A straightforward application of Jensen's inequality for convex functions allows one the retrieve the expected expression $\langle W\rangle\ge\Delta F$. 
The average energetic deviation of a non-equilibrium process from the equivalent reversible isothermal process is known as dissipated work
\begin{eqnarray}
\langle W\rangle_{diss}=\langle W\rangle-\Delta F.
\end{eqnarray}
Due to the Jarzynski equality this quantity is positive, $\langle W\rangle_{diss}\ge 0$. This can be also directly seen from the Crooks relation, taking the logarithm of both sides of the equality in Eq.~\ref{eq:crooks} and integrating over the forward distribution we find
\begin{eqnarray}
\langle \Sigma\rangle=\beta(\langle W\rangle-\Delta F)=K(P_\textrm{F}(W)||P_\textrm{B}(-W))
\end{eqnarray}
 where $K$ is the classical Kullback Leibler divergence and we have introduced the average irreversible entropy change $\langle \Sigma\rangle$ corresponding to the dissipated work. Physically the irreversible entropy change, in this context, would be the internal entropy generated due to the non-equilibrium process which would manifest itself as an additional source of heat if an ideal thermal bath would be reconnected to the system at the end of the protocol. In Ref.~\cite{Deffner2010a} it was shown that the irreversible entropy change can also be expressed in terms of a quantum relative entropy
 \begin{eqnarray}
 \langle\Sigma\rangle=D(\sigma||\tau(\lambda_f,\beta))
 \end{eqnarray}
where $\sigma=U(t_\textrm{f},t_\textrm{i})\tau(\lambda_i,\beta)U^{\dagger}(t_\textrm{f},t_\textrm{i})$ is the out of equilibrium state at the end of the protocol. This is fully consistent with the open system treatment in \cite{spohn}.

\subsection{Phase estimation schemes for extraction of quantum work and heat statistics}

Surprisingly, perhaps one of the most important contributions that ideas from quantum information have made to this field in statistical mechanics is the experimental acquisition of statistics of work. In the classical setting considerable progress has been made in the experimental extraction of the relevant stochastic thermodynamic distributions to explore and verify the fluctuation theorems~\cite{Seifert2012}. Up until very recently, no such experimental progress had been made for quantum systems. A central issue is the problem of building the quantum work distribution as it requires to make reliable projective energy measurements on to the instantaneous energy eigenbasis of an evolving quantum system ~\cite{Talkner2007,Campisi2011}. It was proposed in Ref.~\cite{Huber2008} that these measurements could be reliably performed on a single trapped ion, an experiment that was recently performed \cite{An2014}. 

Alternatives to the projective method have  been proposed~\cite{Dorner2013,Mazzola2013}, based on \emph{phase estimation schemes}, well known in quantum information and quantum optics \cite{Giovannetti2011}. In these schemes, we couple our system to an ancillary system, and perform tomography on that system. The spirit is very similar to the DQC1 algorithm put forward in Ref.~\cite{Knill1998}. The characteristic function of the work probability distribution (Eq.~\ref{eq:charfun}) can be obtained from the ancilla, and the work statistics are then extracted by Fourier transform. 
The {\it characteristic function} is  defined as 
\begin{eqnarray}
\chi_\textrm{F}(u) = \integral{W}{}{}e^{iuW} P_\textrm{F}(W).
\label{eq:charfun}
\end{eqnarray}
The proposals to measure the characteristic function were first tested in the laboratory only quite recently in a Liquid state NMR setup~\cite{Batalhao2014}. This experiment is the first demonstration of the work fluctuation theorems and extraction of work quantum statistics, and is expected to inspire a new generation of experiments at the quantum level. Another interesting extension of these schemes is to go beyond the closed system paradigm and to study open system dynamics at and beyond the weak coupling limit. The first extensions have been proposed in Refs.~\cite{Campisi2013,Mazzola2014,Goold2014c}. In Ref.~\cite{Silva2014} the proposal outlined in Ref.~\cite{Goold2014c} to measure the statistics of dissipated heat was implemented in order to perform a study of the information to energy conversion in basic quantum logic gates at the fundamental Landauer Limit.

Another interesting suggestion made to access the quantum work statistics is the use the concept of a `positive operator valued measure', or POVM \cite{Roncaglia2014}, a well-known concept within quantum information and quantum optics. A POVM is the most general way to describe a measurement in a quantum system, with the advantage that it can always be seen as  a projective measurement on an enlarged system. In this work the authors show that by introducing an appropriate ancilla that the POVM description allows the work distribution to be efficiently sampled with just a single measurement in time. In this work it was suggested that the algorithm proposed could be used, in combination with the fluctuation theorems, to estimate the free energy of quantum states on a quantum computer. The scheme was recently extended and developed in Ref.~\cite{DeChiara2014} along with a promising implementation using ultra-cold atoms. This would be a promising avenue to explore work statistics in a many-body physics setting where the statistics of work can be shown to have universal behaviour at critical points \cite{Mascarenhas2014}. 
 
\subsection{Fluctuation relations with feedback, measurement and CPTP maps}

The relationship between thermodynamics and the information processing is almost as old as thermodynamics itself and is no where more dramatically manifested than by Maxwell's demon \cite{LeffRex90,Plenio2001,LeffRex02,Maruyama2009, Parrondo2015}. One way of understanding the demon paradox is by viewing the demon as performing feedback control on the thermodynamic system. In this case the framework for stochastic thermodynamics and the fluctuation theorems needs to be expanded. Building upon previous work \cite{Sagawa2008,Sagawa2009}, Sagawa and Ueda have generalised the Jarzynski equality to incorporate the feedback mechanism \cite{Sagawa2010,Sagawa2012} for classical systems. This theoretical breakthrough allowed for an experimental demonstration of information to energy conversion in a system by means of of non-equilibrium feedback of a Brownian particle \cite{Toyabe2010}. These feedback based fluctuation theorems were further modified to incorporate both initial and final correlations \cite{Sagawa2012a}. These works, in particular, highlight the pivotal role played by mutual information in non-equilibrium thermodynamics \cite{Parrondo2015}.

The Sagawa-Ueda relations were generalized  to quantum systems in Ref.~\cite{Morikuni2011}. For reasons of pedagogy we will follow this approach here. In the work of Morikuni and Tasaki an isolated quantum system is considered where an external agent has control of the Hamiltonian parameters. The system is initialised in a canonical state, $\tau(\beta)$, and an initial projective measurement of the energy is made whose outcome is $E^{0}_{i}$. The Hamiltonian is then changed via a fixed protocol and evolves according to the unitary operator $U$. In the next stage a projective measurement is performed with outcomes $j=1, \dots,n$ and described by a set of projection operators $\Pi_{1},\dots,\Pi_{n}$. Now the time evolution is conditioned on the outcome $j$ so the Hamiltonian is changed according to these outcomes. This is the feedback control stage. Finally, one makes a projective measurement of the energy of the final Hamiltonian with outcome $E^j_{k}$. In this setting it is shown that 
\begin{equation}
\langle e^{\beta(W-\Delta F)}\rangle=\gamma,
\label{Eq:sagu1}
\end{equation}
where $W=W_{i,j,k}=E^{0}_{i}-E^{j}_{k}$ is the work and $\Delta F$ is the free energy difference between the initial state and the canonical state corresponding to the final value of the Hamiltonian $H^{j}$. We see that in this feedback controlled scenario a new term enters on the right hand side. A straightforward calculation shows that this term evaluates as $\gamma=\sum_{j}\tr[\Pi_jU^{\dagger}_{j}\tau(\beta)U_{j}\Pi_{j}]$. This $\gamma$ quantity is shown in Refs.~\cite{Sagawa2010,Morikuni2011} to be related to the efficiency of the demon in making use of the information it acquires during the feedback process. When it becomes less than one it provides an example of a failed demon who did not make a good use of the information acquired. On the other hand it can become larger than one indicating that the feedback is working efficiently. Another relation discovered by Sagawa and Ueda and quantized by Morkikuni and Taskaki concerns almost the same protocol as just explained only now classical errors are made in the intermediate measurement stage. Again let the intermediate measurement be described by $\Pi_{1},...,\Pi_{n}$ which yield the result $j$ but the controller misinterprets the result as $j'$ with a certain probability. In this framework another generalised fluctuation theorem can be derived, 
\begin{equation}
\langle e^{\beta(W-\Delta F-I)}\rangle=1
\label{Eq:sagu2}
\end{equation}
where $I$ is the mutual information  between the set of measurement outcomes the demon actually records and what is the true result of the projection. These feedback fluctuation theorems for quantum systems were further generalised to the situation when a memory system is explicitly accounted for in Ref.~\cite{Funo2013} and shed light on the amount of thermodynamic work which can be gained from entanglement. In addition to feedback, fluctuation theorems were investigated under continuous monitoring \cite{Campisi2010,Campisi2011a} and analysed for general measurements  \cite{Watanabe2014,Watanabe2014a}.

A recent series of papers have analysed fluctuation-like relations from the operational viewpoint employing the full machinery of  trace-preserving completely positive maps. In Ref.~\cite{Kafri2012}  the formalism is used to give  an alternative derivation of the Holevo bound \cite{Holevo1998}. In Ref.~\cite{Vedral2012} an information-theoretical Jarzynski equality was derived. It was found that fluctuation relations can be derived if the map generated by the open dynamics obeys the unital condition. This has been connected to the breakdown of micro-reversibility for non-unital quantum channels \cite{Rastegin2013,Albash2013,Rastegin2014,Goold2014}. In Ref.~\cite{Goold2015} the authors analysed the statistics of heat dissipated in a general protocol and found that the approach can be used to derive a lower bound on the heat dissipated for non-unital channels. Recently this bound has been used to investigate the connection with the build up of multipartite correlations in collisional models \cite{Lorenzo2015}.

\subsection{Entropy production, relative entropy and  correlations}

With the surge of interest in the thermodynamics of quantum systems and the development of quantum fluctuation relations, research has been directed to microscopic expressions for entropy production. In formulating thermodynamics for non-equilibrium quantum systems, the relative entropy plays a central role~\cite{Sagawa2012}. As first pointed out in Ref.~\cite{Donald1987} this is due to its close relationship with the free energy of a quantum state. The relative entropy also plays a central role in quantum information theory, in particular, in the geometric picture of entanglement and general quantum and classical correlations \cite{Vedral2002,Modi2010}. In the non-equilibrium formulation of thermodynamics  \cite{Campisi2011} it is omnipresent for the description of irreversible entropy production in both closed \cite{Deffner2010a} and open driven quantum systems \cite{Deffner2011} (see also \cite{plastina}). One may then wonder if there exists a relationship between the entropy produced by operations that generate or delete correlations in a quantum state and the measures for correlations in that state? Given the youthful nature of the field the question is largely unanswered but some progress in this direction has been made. 

The relationship between the relative entropy of entanglement and the dissipated work was first proposed as an entanglement witness in Ref.~\cite{Hide2010}. Going beyond the geometric approach a functional relationship between the entanglement generated in a chain of oscillators and the work dissipated was explored in Ref.~\cite{Galve2009} and also later for more general quantum correlations \cite{Carlisle2014}. In an open systems framework it was shown that the irreversible entropy production maybe attributed to the total correlations between the system and the reservoir \cite{Esposito2010} (we note that this derivation is entirely analogous to the formulation of the Landauer principle put forward by in Ref.~\cite{Reeb2014}). The exchange fluctuation relation and the consequences for correlated quantum systems were studied in Ref.~\cite{Jennings2012}.

\subsection{Outlook}
As fluctuation theorems are exact results, valid for arbitrary non-equilibrium dynamics,  they are currently being used to understand the non-linear transport of energy, heat and even information in quantum technologies.  This is a relatively new research avenue and the applications of quantum fluctuation theorems in other fields such as condensed matter physics, quantum optics and quantum information theory are in their infancy. Ultimately, the hope would be that they provide a unifying framework to understand the relationship between information and energy in non-equilibrium quantum systems. Ultimately one would like to form a picture of information thermodynamics of quantum systems under general non-equilibrium conditions.  

As we have seen above, quantum phase estimation, a central protocol in quantum information theory, has been applied successfully to extract work statistics from a small non-equilibrium quantum system and perhaps other such unexpected interdisciplinary links will emerge. For example one wonders if existing experimental schemes could be modified to deal with situations dealing with non-passive initial states so as to study maximal work extraction problems and also to extend to more complicated many-body and open system scenarios. 

In Refs.~\cite{Halpern2014,Dahlsten2015,Salek2015} the first steps towards unification of the work statistics and fluctuation theorems approach to thermodynamics and the single shot statistical mechanics approaches mentioned have been taken (see Sec.~\ref{sec:resource_theories}). We are confident that other links will emerge between various approaches in the not so distant future.

\section{Quantum Thermal Machines}\label{sec:thermal_machines}
In this final brief section of the review we end by considering the area of quantum thermodynamics concerning quantum thermal machines, that is quantum versions of heat engines or refrigerators. We shall overview the extent to which quantum entanglement and correlations are relevant to their operation. 

Whereas in almost all of the above the situation comprised of only one thermal bath and systems in contact with it, in this section our interest is in situations involving two (or more) thermal baths.  Now, there are two regimes which one can focus on: the primary one is usually the cyclic behaviour of systems interacting with the baths, or alternatively the steady state behaviour that is characterised by the currents of heat or work that can be maintained in the long time limit. The second regime is the transient one, and how the system reaches stationarity. 

One way to think of the present situation is that the second thermal bath is the system out of equilibrium with respect to the first bath, and the goal is to produce resources (work, or a steady state current out of a cold bath) at optimal rates. From this perspective, the quantum machine plays the role of the `bridge' or the `mediator' which facilitates the operation of the larger thermal machine. 

The history of quantum thermal machines is a long one, going back to the sixties with the invention of the maser, which can be seen as a heat engine \cite{GeuSchSco67}, and  received much attention over the following decades. A complete overview of the literature in this direction is far beyond the scope of the present review; however excellent recent overviews can be found in Refs.~\cite{Kosloff2013a,Kosloff2014,Gelbwaser2015a}. In the present context, one important message from this body of work is that thermal machines comprised of as little as a single qutrit (3 level system), or of 2 or 3 qubits, can be constructed, that moreover can approach Carnot efficiency (the maximal possible efficiency of any machine). It is thus plausible that they may ultimately become important from the perspective of nanotechnology and implementations of quantum information processes devices. As such a full understanding of their quantum behaviour, including the correlations they can build up, is important. 
Here we review specifically those studies concerned with the role of entanglement and quantum coherence in the functioning of such small quantum thermal machines, both at the level of the machine, as well as in the bath, if pre-processing operations are allowed. We also look at the role of coherence in the transient behaviour when the refrigerator is first switched on. We review a recent proposal for a witness that quantum machines are provably outperforming their classical counterparts. Finally, we look at the idea of using thermal machines as a means of entanglement generation (switching the focus away from the traditions resources of work or heat currents). 

A related idea is that of \emph{algorithmic cooling}, which we summarise in Example 5, and which was recently reviewed in \cite{Park2015}. 

\subsection{Absorption refrigerators}
The first machine we shall look at a quantum model of an absorption refrigerator, a refrigerator which is not run by a supply of external work (which is the situation most customarily considered), but rather run by a source of heat. An absorption refrigerator is thus a device connected to three thermal reservoirs; a `cold' reservoir at temperature $\beta_\rc$ from which heat will be extracted; 
a `hot' reservoir at inverse temperature $\beta_\rh$, which provides the supply of energy into the machine; and finally a `room temperature' reservoir at temperature $\beta_\rr$ into which heat (and entropy) will be discarded. The goal is to cool down the cold reservoir (i.e. extract heat from it). 

There are a number of different figures of merit that one can consider to quantify the performance of the machine. The most commonly considered is the \textit{coefficient of performance} $\COP = Q_\rc/Q_\rh$, where $Q_\rc$ and $Q_\rh$ are respectively the heat currents flowing out of the cold the hot reservoirs (the COP is the analogous quantity to the efficiency for an absorption refrigerator; since the COP can be larger than 1 it cannot be thought of directly as an efficiency). The famous result of Carnot \cite{carnot}, a statement of the second law of thermodynamics, is that the efficiency (or COP) of all thermal machines is bounded as a function of the reservoir temperatures. In particular, for the specific case of an absorption refrigerator we have 
$\COP \leq (\beta_\rr - \beta_\rh)/(\beta_\rc - \beta_\rr)$. 
Other relevant figures of merit are the power $Q_\rc$ (i.e. neglecting how efficient the process is), the COP when running at maximal power, and the minimal attainable stationary temperature $\beta_\rc^{\st}$ for a cold object in contact with the bath.

Below we give a brief outline of the model under consideration, full details of which can be found in Refs.~\cite{Brunner2013,Correa2013a}.  
Consider three qubits, each one in thermal contact with one of the three thermal baths, with local Hamiltonians $H_i = E_i \ket{1}\bra{1}$, for $i = \rc, \rr, \rh$ chosen such that $E_\rr = E_\rc + E_\rh$ to ensure that the system has a degenerate subspace of energy $E_\rr$ formed by the states $\ket{010}$ and $\ket{101}$ (where we use the order $\rc$-$\rr$-$\rh$ for the three qubits). In this subspace the interaction Hamiltonian $H_\rint =g( \ket{010}\bra{101} + \ket{101}\bra{010}) $ is placed, which mediates the transfer of energy. A schematic representation of this fridge can be found in Fig.~\ref{fig:schematic}. 

\begin{bigexample}{Algorithmic cooling}{algo-cooling}
Consider a collection of $n$ qubits, all at inverse temperature $\beta$, with corresponding populations in the ground and excited states $p$ and $(1-p)$ respectively. The goal of algorithmic cooling is to bring $m$ qubits to the ground state by an arbitrary unitary transformation. A fundamental upper bound can be placed on $m$, purely by entropic considerations. The initial entropy is $ nS(\tau(\beta) = n H(p)$, where $H(p) = - p \log_2 p - (1-p) \log_2 (1-p)$ is the \emph{binary Shannon entropy}. Since unitary transformations do not change the entropy, this easily leads to the upper bound on $m$,
\begin{equation}
m \leq n(1-H(p))
\end{equation}
which would be achieved if the remaining $n-m$ qubits are all left at infinite temperature (maximally mixed state) with entropy $S(\tau(0)) = 1$. In \cite{SchVaz99} it was shown that as $n$ tends to infinity this fundamental limit can be approached using an algorithm which uses $O(n\log_2 n)$ unitary gate operations. It was later realised that given access to an external bath this limit can be surpassed: the qubits which end this protocol at infinite temperature can be 'refreshed' to temperature $\beta$ and the protocol can be run again on the remaining $(n-m)$ qubits, for example \cite{Boykin2002}. This is referred to as \emph{heat-bath algorithmic cooling}.  

In order to understand the basic principle, one can focus instead on 3 qubits and assume that the first is the one which is to be cooled down (now not to zero temperature, but any colder temperature). Let us consider the populations of the two states $\ket{100}$ and $\ket{011}$, which are $p^2(1-p)$ and $p(1-p)^2$ respectively. The state$\ket{100}$,  in which qubit one is excited (and therefore `hot') has more population than the state $\ket{011}$, where qubit one is in the ground state (and therefore `cold'). Thus, by swapping the population of these two states the first qubit is cooled down. Indeed, after the application of such a unitary, the final population $p'$ in the ground state of the first qubit is
\begin{equation}
p' = p + (2p-1)p(1-p)
\end{equation}
which is greater than $p$ whenever $(2p-1) > 0$, i.e. whenever the first qubit was at a positive temperature. Finally, a unitary which implements $\ket{011} \leftrightarrow \ket{100}$ whilst leaving all other energy eigenstates the same can easily be constructed from the CNOT and Toffoli gates as 
\[ \Qcircuit @C=1em @R=.7em {
  & \ctrl{1} & \ctrl{2} & \targ & \ctrl{2} & \ctrl{1} & \qw \\
   & \targ & \qw & \ctrl{-1} & \qw & \targ & \qw \\
   & \qw & \targ & \ctrl{-2} & \targ & \qw & \qw
}\]
A recent review giving many more details about algorithmic cooling can be found in Ref.~\cite{Park2015}
\end{bigexample}

\subsubsection{Stationary behaviour}
Assuming the weak coupling regime between the qubits and the baths, the dynamics can be modelled using a time-independent Lindblad Master equation $\dot{\rho} = \mathcal{L}(\rho)$ (with $\mathcal{L}$ the Linbladian, i.e. the most general generator of time-homogeneous, Markovian dynamics). 
The stationary solution $\rho^\rst$, satisfying $ \mathcal{L}(\rho^\rst) = 0$, can be shown to correspond to an absorption refrigerator if the parameters are chosen appropriately, i.e. such that $\beta^\rst_\rc > \beta_\rc$, where $\beta^\rst_\rc$ is the stationary inverse temperature of the cold qubit.

From the point of view of quantum information, the basic questions about this steady state are (i) whether quantum correlations (for example entanglement) are present in the stationary state, and (ii) if yes, whether they are important for the operation, or merely a by-product of quantum evolution. These questions were addressed in Refs.~\cite{Correa2013a,Brunner2013}.

In Ref.~\cite{Correa2013a} quantum correlations in the form of discord were studied. The quantum discord $\mathcal{D}(AB)_\rho := I(A:B)_\rho - I(A:B)_\sigma$, with $\sigma$ the state after a minimally disturbing measurement on Bob, is a form of quantum correlation weaker than entanglement \cite{Ollivier2001b,Henderson2001b}. 
The authors studied quantum discord between numerous inequivalent partitions of the system. The most interesting results were obtained when the discord is calculated between the cold qubit (the qubit which is being cooled) and the relevant subspace of the two remaining qubits (that singled out by the interaction Hamiltonian $H_\rint$). They found that discord is always present, but they found no relationship between the \emph{amount} of discord present and the rate at which heat was extracted from the cold bath. Specifically, to obtain this result they studied the behaviour of discord as a function of the energy spacing $E_\rc$ of the cold qubit. Whilst both quantities typically exhibited local maxima as $E_\rc$ was varied,  these maxima failed to coincide. 

\begin{boxfigure}{Three-qubit fridge }{schematic}
\includegraphics[width=0.4\textwidth]{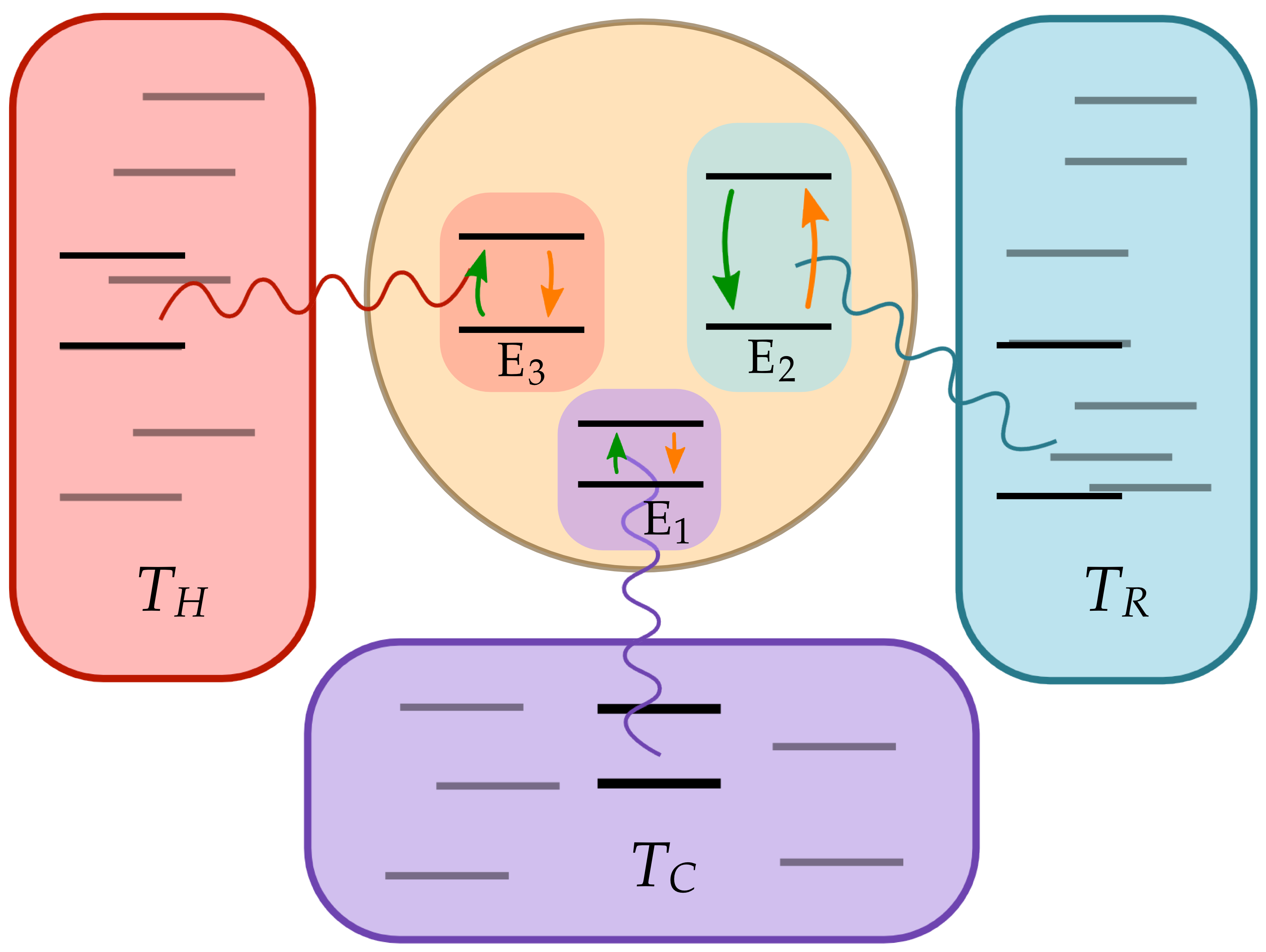}
\tcblower
 Schematic diagram of a three qubit autonomous refrigerator (inside circle), coupled to three thermal reservoirs. The interaction Hamiltonian is represented by the green and orange arrows.
\end{boxfigure}

In Ref.~\cite{Brunner2013} the focus was instead on the entanglement maintained in the steady state. First, if the machine is operating close to the maximal Carnot limit then the state is necessarily fully separable, i.e. a convex combination of product states of the three qubits. 
Conversely, operating far from this regime every type of multipartite entanglement can be found in the stationary state. In particular, there are regimes where entanglement is generated across any fixed bipartition, and even genuine multipartite entanglement can be found, demonstrating that the state has no biseparable decomposition. Here it must be stressed that the amount of entanglement found was small, but that this should be expected due to the weak inter-qubit coupling. 

Finally, it was also shown that there appears to be a link between the amount of entanglement generated in the partition $\rr|\rc\rh$ and the so-called
\emph{cooling advantage} that entangled machines have compared to separable ones. 
In particular, the cooling advantage was defined as the difference between the minimal possible temperatures that could be achieved with either separable or entangled refrigerators. More precisely, by optimising the stationary temperature $\beta_{\rc}^{\st}$ of the cold qubit, varying the Hamiltonian of the machine qubits and their couplings to baths (at fixed temperatures). It was shown that arbitrary machines (i.e. ones allowed to be entangled) could outperform ones which were additionally constrained to be separable. Moreover, the advantage was found to be a function only of the amount of entanglement generated across the $\rr|\rc\rh$ partition. One point of interest is that this is the bipartition of energy entering vs. energy leaving the machine, thus suggesting a connection between the transport properties of the machine and the entanglement. 

\subsubsection{Transient behaviour}
Instead of looking at the steady state behaviour, one may also consider the transient behaviour. Such questions are relevant when one is interested in running a small number of cooling cycles in order to cool down the system as fast as possible. Alternatively, if one is thinking of initialising a system for some other use, the transient regime might also be of interest for quicker initialisation. Intuitively, since the evolution between the qubits is coherent, one might expect the local populations to undergo Rabi oscillations, and hence by running for precise times lower temperatures may be achievable in a transient regime (as the qubits continuous cool down and heat up). 

This is precisely what was shown in Ref.~\cite{Mitchison2015a,Brask2015}. More precisely, in Ref.~\cite{Brask2015} the authors study the Markovian dynamics with weak inter-qubit coupling $g$ (relative to the relaxation rates, as in the above subsection), while in Ref.~\cite{Mitchison2015a} the authors considered additionally Markovian dynamics with strong inter-qubit coupling, and band-limited non-Markovian baths (modelled with a one-qubit memory for each machine qubit). Taking as the natural initial state the product state with qubit to be initially at the same temperature of the bath, both numerically study the transient behaviour of the temperature of the cold qubit as the system approaches stationarity. While in the weak interaction case no Rabi oscillations are observed (since the system is effectively over-damped), in the strong-interaction case Rabi oscillations indeed take place, with period approximately $2\pi/g$. This demonstrates that coherent oscillations offer an advantage for cooling. A more complicated behaviour due to memory effects is also observed in the non-Markovian case in \cite{Mitchison2015a}, but nevertheless the system can be seen to pass through much colder temperatures during its transient behaviour. In Ref.~\cite{Brask2015} it was also shown that if the couplings are chosen appropriately, (in particular such that the weakest coupling is to the hot reservoir), then the system can quickly remains for a long time in a temperature below the stationary temperature, in particular without oscillating above it. This demonstrates a particular stable regime for the preparation of the system at temperatures below its stationary temperature. 

In order to explore more the advantage offered by coherence, Ref.~\cite{Mitchison2015a} also considered varying the initial state, by altering the coherence in the subspace where the Hamiltonian operates. Interestingly, with only a small amount of initial coherence even when considering case (a) of weak-interaction dynamics, oscillations in the temperature are seen, again allowing for cooling below the stationary temperature. In the other two cases, the magnitude of the oscillations is also seen to increase (i.e. the system achieves lower temperatures transiently), demonstrating an advantage in all situations. 

Finally, in Ref.~\cite{Brask2015} the amount of entanglement that is generated in the transient regime was also studied. Focusing on either genuine multipartite entanglement, or entanglement across the partition $\rr|\rc\rh$, i.e. the one corresponding to energy-in vs. energy out (as studied in Ref.~\cite{Brunner2013}), considerably more entanglement can be generated in the transient regime. 

\subsection{Reservoir engineering}
As we have seen in previous sections of the review, thermals states are naturally considered as a free resource which can be utilised and manipulated. Likewise, the ubiquity of thermal machines is that having access to two large thermal reservoirs can also be considered as something essentially free, and thermal machines consider   ways of utilising these resources. 

One interesting avenue is to consider that any transformation of a thermal reservoir which can `easily' be carried out can also be considered to be free, as an idealisation, and this motivates the idea of considering thermal machines which run between \emph{engineered reservoirs}, assuming that the engineering was an easy to perform transformation. In the present context, when one has sufficient control over (part of) the reservoir, then the engineering can be at the quantum level. Here again we are interested specifically in the role that quantum correlations engineered in the bath have on the functioning of quantum thermal machines.

In Refs.~\cite{Roßnagel2014a,Correa2014d} reservoir engineering in the form of \emph{squeezing} is considered, since squeezing is relatively easy to carry out, and is furthermore known to offer quantum advantages in other contexts in quantum information. That is the reservoir, instead of consisting of a large collection of modes in thermal states at inverse temperature $\beta_\rh$, are in fact squeezed thermal states (at the same temperature). More precisely, the squeezing operator is $U_\mathrm{sq} = \exp((ra^2 - r^*{a^\dagger}^2)/2)$ with $a$ and $a^\dagger$ the annihilation and creation operators respectively, and the squeezed thermal state (of a given mode, i.e. a harmonic oscillator) is $U_\mathrm{sq}\tau(\beta)U_\mathrm{sq}^\dagger$. Whereas normally the variances of the quadratures ($x = (a + a^\dagger)/2$ and $p = (a - a^\dagger)/2i$) are symmetric, the squeezed modes become asymmetric, with one the former amplified by the factor $e^{r}$, and the latter shrunk by $e^{-r}$. The important point is that a system placed in thermal contact with such a squeezed reservoir will not thermalize towards a thermal state at $\beta$, but rather to a squeezed thermal state, which has the same average number of photons as a thermal state at temperature $\beta(r) < \beta$. That is, in terms of average number of photons, a squeezed thermal state appears `hotter' than a thermal bath. 

Starting first with Ref.~\cite{Correa2014d}, a model of an absorption refrigerators is considered, identical to the one outlined in the previous section. Here, in accordance with the above, in the weak coupling regime the effect of the reservoir engineering amounts to modifying the Linbladian $\mathcal{L}$, such that the term corresponding to the hot reservoir $\mathcal{L}_\rh$ transforms to $\mathcal{L}_{\rh}(r)$, where this now generates dissipation towards the squeezed thermal state at $\beta_\rh(r)$ . 
They show that maximal COP that the refrigerator can approach becomes 
\begin{equation}
\eta(r) = \frac{\beta_\rr - \beta_\rh(r)}{\beta_\rc - \beta_\rr} > \eta_\mathrm{c} = \frac{\beta_\rr - \beta_\rh}{\beta_\rc - \beta_\rr}.
\end{equation}
That is, the COP overcomes the Carnot limit that bounds the COP of any absorption refrigerator operating between baths at $\beta_\rc$, $\beta_\rr$ and $\beta_\rh$, if reservoir engineering is not carried out. Thus if reservoir engineering is more readily available than a hotter `hot' bath, then this approach clearly provides an advantage in terms of COP. 

In Ref.~\cite{Roßnagel2014a} a different model was considered, this time a quantum heat engine operating a quantum Otto cycle, a time dependent cycle, comprising two expansion stages (changing the Hamiltonian of the system) and two thermalization stages. This system considered comprised of a single harmonic oscillator, with initial spacing $E_1$. While uncoupled to any environment, the first stage is an expansion, whereby $E_1 \to E_2 > E_1$, i.e. the Hamiltonian is changed in time. In the second stage the system is then placed in contact with a squeezed hot reservoir (this is the stage which differs from a standard Otto cycle, where an unsqueezed hot reservoir is used). After disconnection, the third stage is a compression stage, bringing the spacing back to from $E_2$ to $E_1$. Finally, the system is placed in contact with a cold (unsqueezed) reservoir, in order to thermalize at the cold temperature. This cycle is summarised in Fig. \ref{fig:schematicOtto}. 
The authors perform an analysis of the system and similarly show that the maximum efficiency of the engine exceeds the Carnot efficiency (of the Otto cycle, $\eta = \beta_\rh/\beta_\rc$). Moreover, if one considers the efficiency at maximum power, then this can also be surpassed, and as the squeezing parameter becomes large, the efficiency at maximum power approaches unity.

\begin{boxfigure}{Quantum Otto Engine}{schematicOtto}
\includegraphics[width=0.4\textwidth]{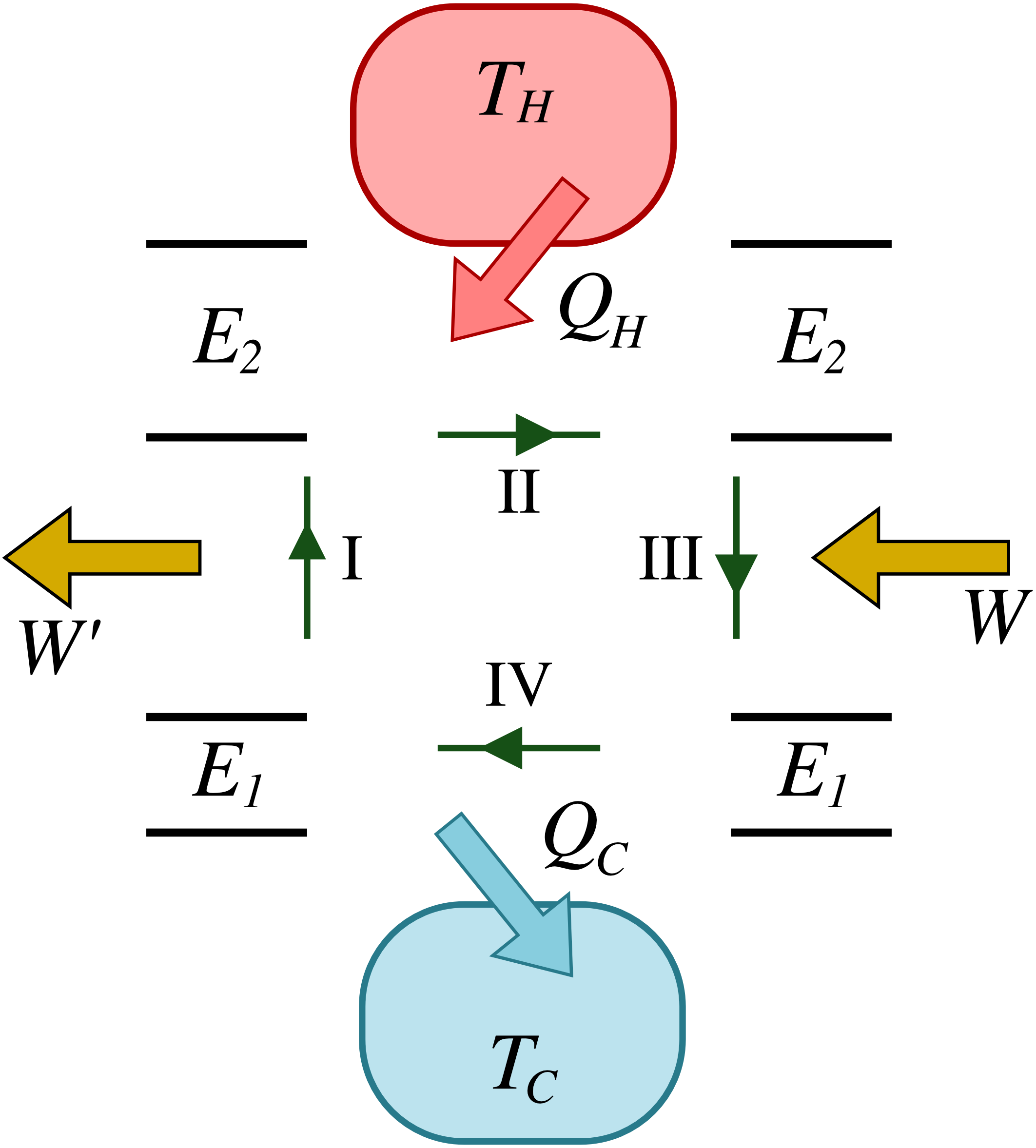}
\tcblower
 Schematic diagram of a quantum Otto engine, depicting only a single pair of levels, which can represent either a qubit, or the spacing of a harmonic oscillator. The four stages are (I) expansion from $E_1$ to $E_2$. (II) contact with a hot thermal reservoir. (III) compression from $E_2$ to $E_1$. (IV) contact with a cold thermal reservoir. 
\end{boxfigure}

Finally, we stress that these results do not constitute a violation of the second law, since they consider a scenario outside the regime of applicability of the Carnot limit (much in the same way that a regular car engine, consuming fuel, does not violate the second law, since it is also outside the regime of applicability). Conversely, it is interesting that the net effect of squeezing appears to be \textit{as if} the hot reservoir has been heated to a temperature $\beta_\rh(r)$, and that the performance of the machines is bounded exactly by the Carnot limit with respect to this new temperature.

\subsection{Quantum thermodynamic signatures}
One way to differentiate between a system which is genuinely using quantum effects and one which is only using the formal structure of quantum mechanics (the discreteness of energy levels, for example) is to devise signatures, or witnesses, for quantum behaviour. This is similar to what is done in entanglement theory, or in Bell nonlocality, where one finds witnesses which certify that entanglement was present, since no separable quantum state could pass a certain test. An interesting question is whether one can find analogous witnesses in a quantum thermodynamics setting. This is what was proposed in Ref.~\cite{Uzdin2015a} in the form of \emph{Quantum thermo signatures}.

In more detail, the main idea of Ref.~\cite{Uzdin2015a} is to find a threshold on the power of a thermal machine which would be impossible to achieve for a machine which is `classical'. The authors take as the minimal set of requirements for a machine to be considered classical (i) that it's operation can be fully described using population dynamics (i.e. as a rate equation among the populations in the energy eigenbasis); (ii) that the energy level structure and coupling strengths are unaltered compared to quantum model under comparison; (iii) that no new sources of heat or work are introduced. A way to satisfy the above three constraints is to add pure de-phasing noise in the energy eigenbasis on top of the dissipative dynamics of the quantum model (arising from the interaction with the thermal reservoirs). One can then compare models with and without de-phasing noise, and ask whether the additional noise places an upper bound on the power of the machine.

For simplicity in presentation, in what follows we will focus here on the results obtained for the four-stage qubit Otto heat engine, similar to the one described in the previous subsection (except now with a qubit in place of a harmonic oscillator).  We note that the authors show that the same results hold for a two-stage engine \cite{Allahverdyan2010} and for continuous time engines \cite{Scovil1959}, as well as for refrigerators and heat pumps. As an aside, the reason why the result holds for all three models is because Ref.~\cite{Uzdin2015a} also proves that in the regime of weak-coupling to the bath, and weak driving, all three types of engine can be shown to be formally equivalent, producing the same transient and steady state behaviour at the level of individual cycles. 

It is shown that a state independent bound can be placed on the power of a classical machine which is proportional to the duration of a single cycle of the engine $\tau_\mathrm{cyc}$, as long as the so-called `engine-action' $s$ is small, where the engine action is the product of the duration $\tau$ and energy scale (as measured by the operator norm of each term appearing in the Master equation). They demonstrate that there is a regime where a quantum engine (i.e. one without additional dephasing) can provably outperform the corresponding classical machine, with powers an order of magnitude larger in the former case.

\subsection{Stationary entanglement}
Entanglement is understood to be a fragile property of quantum states, that is one typically expects that noise will destroy the entanglement in a quantum state. Much effort has been invested in investigating and devising ways in which one can counter the effects of noise, and maintain entanglement in a system, such as quantum error correction, dynamical decoupling, decoherence free subspaces, to name but a few. 

In the first subsection we saw that the non-equilibrium steady state of autonomous quantum thermal machines can be entangled. If one thus focuses not on their thermodynamic functioning, but rather on their entanglement functioning, we see that whenever a thermal machine reaches a steady state which is entangled, this constitutes a way of generating stead state entanglement, merely through dissipative interactions with a number of thermal environments at differing temperatures. 

Furthermore, if the interest is only in steady state entanglement generation, then it is not even necessary that the machine perform any standard thermodynamic task, and can in fact simply be a \emph{bridge} between two reservoirs, allowing the steady flow of heat from hot to cold such that the stationary state of the bridge is necessarily entangled. This is precisely the situation which was first considered in Ref.~\cite{Quiroga2007a}, where the minimal system of two qubits interacting with two baths at temperatures $\beta_\rh$ and $\beta_\rc$ was considered in the weak coupling (Markovian) regime. Numerous variants were then discussed: in Refs.~\cite{Sinaysky2008b,Ban2009b,Ferraro2010,Kheirandish2010,Scala2010} different aspects of the dynamical approach to the steady state were analysed (assuming non-Markovian dynamics, the rotating wave approximation, etc); in Refs.~\cite{Huang2009b,Pumulo2011} a 3 qubit bridge was considered; in Ref.~\cite{Wu2011a} the stationary discord was also studied; in Ref.~\cite{Bellomo2013d,Bellomo2013a} geometric and dielectric properties of the environment were considered, and in Ref.~\cite{Brask2015a} superconducting flux qubits and semiconductor double quantum dot implementations were explored. 

Focusing on the simplest possible example, that of the two qubit bridge, the take home message of this line of investigation is that this is a viable means to generating stationary entanglement. In particular the implementations considered in Ref.~\cite{Brask2015a} suggest that in experimentally accessible situations steady state entanglement can indeed be maintained at a level which might be usable to then later distill. 



\subsection{Outlook}
We have seen in this section a range of results concerning quantum thermal machines, focusing primarily on the quantum correlations and entanglement present in the machine, as well as other signatures of quantumness. Although we have focused on the progress that has been achieved so far, there are a number of directions which should be explored in further work to more fully understand the role of quantum information for quantum thermal machines.

First of all, the main playground of study in this section has been the weak-coupling regime, where the machine is in weak thermal contact with the thermal reservoirs. It is important and interesting to ask what happens outside of this regime, when the thermal baths are strongly coupled to the machine. On the one hand, intuition suggests that stronger coupling corresponds to more noise, which will be detrimental to fragile quantum correlations. On the other hand, stronger driving might lead to more pronounced effects. As such, the interplay between noise and driving needs to be better understood.

Second, we have seen that quantum signatures, either in terms of entanglement or coherence, can be constructed, which show that there is more to quantum thermal machines than just the discreteness of the energy levels. Here, it would be advantageous to have more examples of quantum signatures, applicable in as wide a range of scenarios as possible. An experimental demonstration of a quantum signature would also be a great development concerning the implementation of thermal machines.

Finally, thinking of cooling as a form of error correction, it is interesting to know if ideas from quantum thermal machines can be incorporated directly into quantum technologies as a way to fight de coherence. This would be as an alternative to standard quantum error correction ideas, and an understanding of how they fit alongside each other could be beneficial from both perspectives.

\section{Final Remarks}

Ideas coming from quantum information theory have helped us understand questions, both fundamental and applied, about the thermodynamic behaviour of systems operating at and below the verge at which quantum effects begin to proliferate. 
In this review we have given an overview of these insights. We have seen that they have been both in the form of technical contributions, for example with new mathematical tools for old problems, such as the equilibration problem, and also in the form of conceptual contributions, like the resource theory approach to quantum thermodynamics. 

Although quantum information is only one of the many fields currently contributing to quantum thermodynamics, we expect its role  to become more important as the field grows and matures. Indeed, we believe that placing information as a central concept, just as Maxwell did when his demon was born, will lead to a deeper understanding of many active areas of physics research beyond quantum thermodynamics.


\section*{Author contributions}
All authors contributed equally to this review. Sections I and III were adapted from LdR's PhD thesis~\cite{DelRio2015}.

\section*{Acknowledgements}
We thank Fernando Brandao, Aharon Brodutch, Nicolai Friis, Marti Perarnau-Llobet, Joe Renes, Raam Uzdin and Nicole Yunger-Halpern for helpful feedback on the manuscript. LdR thanks support from ERC AdG NLST and EPSRC grant DIQIP. MH acknowledges funding from the Juan de la Cierva fellowship (JCI 2012-14155), the European Commission (STREP `RAQUEL') and the Spanish MINECO Project No. FIS2013-40627-P, the Generalitat de Catalunya CIRIT Project No. 2014 SGR 966. MH furthermore acknowledges funding through the AMBIZIONE grant PZ00P2\_161351 from the Swiss National Science Foundation (SNF). PS Acknowledges support from the European Union (Projects FP7-PEOPLE-2010-COFUND No. 267229, ERC CoG QITBOX and ERC AdG NLST). AR thanks support from the Beatriu de Pinos fellowship (BP-DGR 2013), the EU (SIQS), the Spanish Ministry Project FOQUS (FIS2013-46768-P), the Generalitat de Catalunya (SGR 874 and 875) and the Spanish MINECO (Severo Ochoa grant SEV-2015-0522). All authors acknowledge the COST Action MP1209.

\bibliography{thermo}

\end{document}